\documentclass[aps,pre,twocolumn,pacs]{revtex4-1}

\usepackage{bm}
\usepackage{amsmath}
 \usepackage{amssymb}
 \usepackage{graphicx}
\usepackage{subfigure}

\begin{document}

\title{Confluent and non-confluent phases in a model of cell tissue}
\author{Eial Teomy and David A. Kessler,$^*$ }
\affiliation{Department of Physics, Bar-Ilan University, Ramat-Gan 52900, Israel}

\author{Herbert Levine }
\affiliation{Center for Theoretical Biological Physics, Rice University, Houston, TX 77005, U.S.A.}
\date{\today}

\begin{abstract}
The Voronoi-based cellular model is highly successful in describing the motion of two-dimensional confluent cell tissues. In the homogeneous version of this model, the energy of each cell is determined solely by its geometric shape and size, and the interaction between adjacent cells is a byproduct of this additive energy. We generalize this model so as to allow zero or partial contact between cells. We identify several phases, two of which (solid confluent and liquid confluent) were found in previous studies that imposed confluency but others that are novel. Transitions in this model may be relevant for understanding both normal development as well as cancer metastasis. 
\end{abstract}

\maketitle

\section{Introduction}
In many biological processes, epithelial cells adopt collective organization, stabilized by cell-cell adhesions. The simplest possibility for such a system is a confluent phase where there is no free space between any of the cells, but other types of organization are also possible. For example, individual cells or small cell clusters can, under some conditions, undergo phenotypic transitions and break free of the bulk confluent tissue; this could be relevant for processes such as cancer metastasis~\cite{weinberg-review}. It is clearly of interest to develop a model which can identify under what conditions the system exhibits nonconfluent behavior and what different nonconfluent phases are possible.

In this paper, we consider Voronoi-based models \cite{Brodland2002,Brodland2004,Pathmanathan2009,Graner1993a,Graner1993b,Schaller2005,Bi2016,Barton2017} (which are closely related to vertex models \cite{Honda1980,Weliky1990,Nagai2001,Fletcher2014,Fletcher2013,Brodland2004,Bi2015}) which have often been used to simulate cell tissue. In these models, the shapes of cells in a confluent aggregate is approximated as a convex polygon. Although such an approximation cannot capture all morphologies, it is sufficient in many cases. Most previous studies considered confluent tissues in either periodic \cite{Nagai2001} or open boundary conditions \cite{Honda1980,Pathmanathan2009,Barton2017}. Although periodic boundary conditions may well describe the behavior of cells inside a large cluster, they cannot address the behavior of cells at the boundary of the cluster.  Furthermore, in order to account for separation of cells from the main cluster, one needs to go beyond models which focus only on confluent tissues.  There have in fact been several works which have allowed for non-confluent cell arrangements \cite{Graner1993a,Graner1993b,Schaller2005} in Voronoi-based models. The basic idea is to introduce a length scale, requiring that each cell lie entirely within a distance $\ell$ of the reference point that describes its location. As a result, the cell boundaries may consist not only of polygonal segments, but also circular arcs, and there can be intercellular regions between cells. These works, while introducing this class of model, did not address the possible phases predicted by such a system.

There are several ways to account for the forces acting by and on the cells. Here we consider an energy functional which depends on the shape of the cells, specifically the quadratic energy functional used in \cite{Farhadifar2007}. However, as we argue later, the specific form of the energy functional is not key. Besides these forces, the cells also perform active motion, which we approximate here as ``thermal" fluctuations induced by an effective temperature \cite{Szamel2015,Steffenoni2016}. This, of course, rules out the possibility of organized cell motion, which requires an extension to the model; this will be addressed elsewhere. We first present the model and discuss some basic structures which can serve as buildings blocks for the organization of the tissue. We then present both simulations and analytic estimates of the zero temperature phase diagram and close with  a discussion of finite temperature effects. 

\section{The model}

In Voronoi-based models, each of the $N$ cells is defined by its reference point, $\vec{r}_{i}$, and it contains all the points which are closest to its reference point. This is equivalent to a Voronoi tessellation of the plane. We consider a two-dimensional model and use the quadratic energy functional used by \cite{Brodland2002,Ouchi2003,Farhadifar2007} according to which the total energy of the system is given by
\begin{align}
E=\sum_{i}K_{i}\left(A_{i}-A_{0,i}\right)^{2}+\sum_{i}\Gamma_{i}P^{2}_{i}+\sum_{<i,j>}\Lambda_{i,j}P_{i,j} ,\label{energy0}
\end{align}
where $A_{i}$ and $P_{i}$ are the area and circumference respectively of cell $i$, $P_{i,j}$ is the length of the interface between cells $i$ and $j$, $A_{0,i}$ is the target area of cell $i$, $K_{i}$ is the elastic modulus of cell $i$, $\Gamma_{i}$ is the contractility coefficient of cell $i$, and $\Lambda_{i,j}$ is the tension between cells $i$ and $j$. The first two sums are over all cells, and the third sum is over all pairs of neighboring cells.  As the cells are not necessarily confluent, we also need to consider in the third sum the contribution to the energy of interfaces between cells and unoccupied regions. We consider herein the 
situation where all the cells have the same $K_{i}$, $A_{0,i}$, $\Gamma_{i}$ and $\Lambda_{i,j}$. If, in addition, the tensions for cell-cell interfaces and cell-unoccupied region interfaces are identical, then up to an unimportant additive constant the energy takes the simple form
\begin{align}
E=K_{A}\sum_{i}\left(A_{i}-A_{0}\right)^{2}+K_{P}\sum_{i}\left(P_{i}-P_{0}\right)^{2} .\label{energy1}
\end{align}
Physically, it is more appropriate to consider different tensions for cell-cell interfaces and cell-boundary interfaces. However, this does not affect the phase diagram qualitatively, but merely moves the location of the phase boundaries, as we show in Appendix \ref{dif_inter}. Therefore, for simplicity, we assume in the following that the cell-cell tension is equal to the cell-unoccupied region tension.
On top of these four parameters and the $2N$ degrees of freedom, there is, as noted above, another length scale $\ell$ representing the maximal radius of each cell. $\ell$ can be either fixed as a control parameter, or it can be another degree of freedom, subject to optimization to minimize the global energy. This latter case will be referred to as the dynamical $\ell$ model.

We consider here overdamped dynamics and approximate the active motion of the cells as ``thermal" fluctuations induced by an effective temperature $T$, such that each reference point advances in time according to an overdamped Langevin equation.
\begin{align}
\frac{d\vec{r}_{i}}{dt}=-\frac{1}{\xi}\vec{\nabla}_{i}E+\sqrt{\frac{2k_{B}T}{\xi}}\vec{\eta}_{i}(t) ,\label{eveq}
\end{align}
where $\xi$ is the effective damping strength, $\vec{\nabla}_{i}$ stands for the derivative of the energy with respect to $\vec{r}_{i}$, $k_{B}$ is the Boltzmann constant, $T$ is the effective temperature, and $\vec{\eta}_{i}$ is a white noise with zero mean and
\begin{align}
\langle\vec{\eta}_{i}(t)\vec{\eta}_{j}(t')\rangle = \delta_{i,j}\delta(t-t') .
\end{align}
If in addition $\ell$ is taken to be a free variable which changes so as to minimize the energy, its dynamics are described by $\frac{\partial\ell}{\partial t}=-\zeta\frac{\partial E}{\partial\ell}$ .
Hence, at equilibrium we require $\partial E/\partial\ell=0$, or equivalently
\begin{align}
0=2\left(\langle\tilde{A}^{2}\rangle-\tilde{A}_{0}\langle\tilde{A}\rangle\right)+\tilde{K}\left(\langle\tilde{P}^{2}\rangle-\tilde{P}_{0}\langle\tilde{P}\rangle\right) .\label{zerol}
\end{align} which follows from rescaling the reference points by $\ell$

\subsection{No density-independent phase transition in confluent tissues}
\label{sec_proof}

In previous works on the vertex model and Voronoi-based model under periodic boundary conditions, it was claimed that the system undergoes a density-independent phase transition \cite{Bi2015,Bi2016}. Here we prove that there is no physical density-independent phase transition in confluent tissues. In order to have a density-independent phase transition, the density should be a control parameter. If the number of cells $N$ is fixed and the tissue is confluent, the density is trivially related to the average area $\langle A\rangle$. We therefore parametrize the density by $\ell=\sqrt{\langle A\rangle}$, such that $\langle \tilde{A}\rangle=1$.

First, note that under these conditions all the terms in the energy which depend on $A_{0}$ do not depend on the free variables, but rather only on the control parameters. Thus, these terms may be considered to be unimportant additive constants. Hence, the parameter $A_{0}$ cannot affect any kind of transition. This means that the phase transition investigated in \cite{Bi2015,Bi2016} is not governed by the density-independent dimensionless parameter $P_{0}/\sqrt{A_{0}}$, but rather by $P_{0}/\sqrt{\langle A\rangle}$ which is clearly density-dependent.

In fact, there are no physical density-independent phase transitions under these conditions. Using Landau's theory of phase transitions, we now show that there are no density-independent phase transitions as long as both $K_{A}$ and $K_{P}$ are positive.

First note that the only dimensionless density-independent parameter is
\begin{align}
\hat{K}=\frac{K_{P}}{K_{A}P^{2}_{0}} .
\end{align}
Second, rewrite the energy as
\begin{align}
\hat{E}=\hat{\ell}^{4}\left\langle\tilde{A}^{2}\right\rangle+\hat{K}\left(\hat{\ell}^{2}\left\langle\tilde{P}^{2}\right\rangle-2\hat{\ell}\left\langle\tilde{P}\right\rangle\right) ,\label{energy_density}
\end{align}
with
\begin{align}
&\hat{E}=\frac{E}{NK_{A}P^{4}_{0}} ,\nonumber\\
&\hat{\ell}=\frac{\ell}{P_{0}} .
\end{align}
According to to the Landau theory of phase transitions, near the phase transition the energy density may be approximated by
\begin{align}
\hat{E}=a+r\Psi^{2}+s\Psi^{4}+H\Psi ,
\end{align}
where $\Psi$ is the order parameter controlling the transition. Due to the form of the energy, Eq. (\ref{energy_density}), the $\ell$ dependence of the parameter $r$ is of the form
\begin{align}
r=\hat{\ell}^{4}A^{(2)}_r+\hat{K}\left(\hat{\ell}^{2}P^{(2)}_{r}-2\hat{\ell}P^{(1)}_{r}\right) ,\label{r-shape}
\end{align}
where 
\begin{gather}
A^{(2)}_{r}=\frac{d^2\langle \tilde{A}^2\rangle}{d\Psi^2}\Big|_{\Psi=0},\qquad P^{(2)}_{r}=\frac{d^2\langle \tilde{P}^2\rangle}{d\Psi^2}\Big|_{\Psi=0}\nonumber\\ P^{(1)}_{r}=\frac{d^2\langle \tilde{P}\rangle}{d\Psi^2}\Big|_{\Psi=0} 
\end{gather}
with similar expressions for $a$, $s$ and $H$.  Note that the coefficients $A_r^{(2)}$, $P_r^{(1)}$, $P_r^{(2)}$
 do not depend on $\hat{K}$ and $\hat{\ell}$. In particular that $r$ is in general a fourth order polynomial of $\ell$. If the phase transition depends only on $\hat{K}$, then $r$ has the form
\begin{align}
r=r_{0}\left(\hat{K}-\hat{K}_{c}\right) .\label{r-shape2}
\end{align}
Combining Eqs. (\ref{r-shape}) and (\ref{r-shape2}) we find that
\begin{align}
&r_{0}=\hat{\ell}^{2}P^{(2)}_{r}-2\hat{\ell}P^{(1)}_{r} ,\nonumber\\
&\hat{K}_{c}=-\frac{\hat{\ell}^{4}A^{(2)}_{r}}{\hat{\ell}^{2}P^{(2)}_{r}-2\hat{\ell}P^{(1)}_{r}} .
\end{align}
Since $\hat{\ell}$ scales linearly with $\ell$, while $A^{(2)}_{r}$, $P^{(2)}_{r}$ and $P^{(1)}_{r}$ are independent of $\ell$, the only way for the phase transition to be density-independent is to set $\hat{K}_{c}=0$. However, since $\hat{K}>0$ always, there is no such phase transition. 

If we relax the physical constraint that $\hat{K}>0$, a density-independent phase transition could occur. The effects of such an unphysical phase-transition might be seen as $\hat{K}$ decreases and the system nears the unobtainable critical point. However, we do not know what order parameter undergoes such a transition.

\section{Second-order phase transitions}
\label{sec_per.vs.free}

For a given set of dimensionless parameters $\tilde{P}_{0} = P_0/\ell$, $\tilde{A}_{0} = A_0 /\ell ^2$ and $\tilde{K}= K_p/K_A \ell ^2$, the system reaches a local minimum of the energy in some particular configuration. We are interested in second-order phase transitions, at which this configuration changes to a local maximum or a saddle point of the energy. These transitions can readily occur even in simple configurations. We consider here two configurations: a honeycomb lattice in a periodic system and a rosette configuration in a system with open boundary conditions.

\subsection{Honeycomb lattice}
We first investigate the stability of a honeycomb lattice in an infinite or periodic boundary conditions. Due to the symmetry of the configuration we find that it satisfies
\begin{align}
\vec{\nabla}_{i}\left\langle\tilde{{\cal{O}}}\right\rangle=0 ,\label{dpi0}
\end{align}
for ${\cal{O}}=A,A^{2},P$ and $P^{2}$ where $\vec{\nabla}_{i}$ is the derivative with respect to the location of cell $i$. Thus, the honeycomb lattice is an extremum of the energy.

In order to check that it is in fact a minimun of the energy, and so a stable configuration, we perturb it such that each particle is located at
\begin{align}
\vec{r}_{i}=\vec{r}^{s}_{i}+\vec{\epsilon}_{i} ,
\end{align}
where $\vec{r}^{s}_{i}$ is the location of cell $i$, and all the vectors are normalized by $\ell$. Under the harmonic approximation, the force acting on cell $i$ in direction $\alpha$ is
\begin{align}
F_{i,\alpha}=-\frac{\partial\tilde{E}}{\partial r_{i,\alpha}}=-\sum_{j,\beta}\left.\frac{\partial^{2}\tilde{E}}{\partial r_{i,\alpha}\partial r_{j,\beta}}\right|_{\vec{r}^{s}_{i},\vec{r}^{s}_{j}}\epsilon_{j,\beta} .\label{force1}
\end{align}
Using the Fourier transform of $\vec{\epsilon}_{i}$
\begin{align}
\vec{\epsilon}_{i}=\int d\vec{q}\vec{C}\left(\vec{q}\right)e^{i\vec{q}\cdot\vec{r}_{i}-\omega\left(\vec{q}\right) t} ,
\end{align}
Eq. (\ref{force1}) transforms into
\begin{align}
F_{i,\alpha}=-\int d\vec{q}\sum_{j,\beta}C_{\beta}\left(\vec{q}\right)\left.\frac{\partial^{2}\tilde{E}}{\partial r_{i,\alpha}\partial r_{j,\beta}}\right|_{\vec{r}^{s}_{i},\vec{r}^{s}_{j}}e^{i\vec{q}\cdot\vec{r}_{j}-\omega(\vec{q})t} .\label{force2}
\end{align}
Using the relation $\vec{F}=\xi\vec{v}$ we find that
\begin{align}
&-\int d\vec{q}\sum_{j,\beta}C_{\beta}\left(\vec{q}\right)\left.\frac{\partial^{2}\tilde{E}}{\partial r_{i,\alpha}\partial r_{j,\beta}}\right|_{\vec{r}^{s}_{i},\vec{r}^{s}_{j}}e^{i\vec{q}\cdot\vec{r}_{j}}=\nonumber\\
&-\xi\int d\vec{q}\omega\left(\vec{q}\right)C_{\alpha}\left(\vec{q}\right)e^{i\vec{q}\cdot\vec{r}_{i}} .
\end{align}
The system is stable only if $\omega\left(\vec{q}\right)$ is positive for all $\vec{q}$, or equivalently if the $2\times2$ matrix $\Omega\left(\vec{q}\right)$ have only positive eigenvalues for all $\vec{q}$, where
\begin{align}
\Omega_{\alpha,\beta}=\sum_{j}\left.\frac{\partial^{2}\tilde{E}}{\partial r_{i,\alpha}\partial r_{j,\beta}}\right|_{\vec{r}^{s}_{i},\vec{r}^{s}_{j}}e^{i\vec{q}\cdot\left(\vec{r}_{j}-\vec{r}_{i}\right)} .\label{omegadef}
\end{align}
The full form of the matrix $\Omega$ for a honeycomb structure appears in Appendix \ref{app_honeycomb}.

In order for this configuration to be stable, we require that both eigenvalues of the matrix $\Omega$ are positive for all values of $\vec{q}$. The two eigenvalues of $\Omega$ are
\begin{align}
\omega_{\pm}=\frac{\Omega_{x,x}+\Omega_{y,y}}{2}\pm\sqrt{\left(\frac{\Omega_{x,x}-\Omega_{y,y}}{2}\right)^{2}+\Omega^{2}_{x,y}} .
\end{align}
Hence, the transition occurs when the minimum of $\omega_{-}$ is zero for some $\vec{q}$. Note that this occurs when the minimum of the determinant of $\Omega$,
\begin{align}
|\Omega|=\Omega_{x,x}\Omega_{y,y}-\Omega^{2}_{x,y} ,
\end{align}
is zero for some $\vec{q}$. The determinant of $\Omega$ may be written as
\begin{align}
|\Omega|=&\frac{\tilde{K}\left(\tilde{P}-\tilde{P}_{0}\right)}{18\tilde{P}^{2}}\times\nonumber\\
&\times\left\{\tilde{P}^{3}g_{1}(\vec{q})+23328\tilde{K}\left[\tilde{P}g_{2}(\vec{q})-\tilde{P}_{0}g_{3}(\vec{q})\right]\right\} ,\label{fung}
\end{align}
with the full expression of the functions $g_{i}(\vec{q})$ appearing in Appendix \ref{app_honeycomb}.

Straightforward analysis yields $g_{1}(\vec{q})\geq 0$ and $0\leq g_{3}(\vec{q})\leq g_{2}(\vec{q})\leq 1$ for all $\vec{q}$. Therefore, for $\tilde{P}>\tilde{P}_{0}$ we find that $|\Omega|>0$ for all $\vec{q}$. Hence, for $\tilde{P}>\tilde{P}_{0}$ both eigenvalues $\omega_{\pm}$ have the same sign for all $\vec{q}$. Choosing some arbitrary $\vec{q}$, we find that the sign is positive, and thus the structure is stable for $\tilde{P}>\tilde{P}_{0}$.

In order to show that the structure is unstable for $\tilde{P}<\tilde{P}_{0}$, we need to show that in this range either (1) $|\Omega|<0$ for some $\vec{q}$ or (2) $|\Omega|>0$ and $\omega_{+}<0$ for all $\vec{q}$. We now analyze the cubic polynomial in $\tilde{P}$
\begin{align}
G(\tilde{P})=\tilde{P}^{3}g_{1}(\vec{q})+23328\tilde{K}\left[\tilde{P}g_{2}(\vec{q})-\tilde{P}_{0}g_{3}(\vec{q})\right] . 
\end{align}
Since $g_{1}(\vec{q})\geq0$ and $g_{2}(\vec{q})\geq0$ for all $\vec{q}$, the polynomial $G$ is an increasing function of $\tilde{P}$. Since at $\tilde{P}=0$ the polynomial $G$ is negative, and at $\tilde{P}\rightarrow\infty$ the polynomial $G$ goes to infinity, it must have a single root at some positive $\tilde{P}^{\ast}(\vec{q})$. We now define $\tilde{P}^{\dagger}$ by
\begin{align}
\tilde{P}^{\dagger}=\min_{\vec{q}}\tilde{P}^{\ast}(\vec{q}) ,
\end{align}
such that for $\tilde{P}>\tilde{P}^{\dagger}$ there is some $\vec{q}$ for which $G>0$ and thus $|\Omega|<0$ (as we consider now only the range $\tilde{P}<\tilde{P}_{0}$). In the range $\tilde{P}<\tilde{P}^{\dagger}$, the sign of $\omega_{+}$ is the same for all values of $\tilde{P}$ and $\vec{q}$. Choosing an arbitrary value of $\tilde{P}$ and $\vec{q}$, we find that $\omega_{+}<0$, and thus the structure is unstable for $\tilde{P}<\tilde{P}_{0}$.

In summary, in a honeycomb lattice in an infinite system, $\ell$ plays no role, and the calculation recapitulates that of \cite{Farhadifar2007} and \cite{Staple2010}. The structure becomes unstable  at $\tilde{P}_{0}=q_6 \sqrt{\tilde{A}_0}$, independently of $\tilde{K}$, where $q_6=\sqrt{24/\tan(\pi/6)}\approx 3.7$. For the dynamical $\ell$ model, we additionally impose Eq. (\ref{zerol}), and find that at the critical point the mean area satisfies $\langle A \rangle={A}_{0}$.

\subsection{Rosette Configuration}

A second example is provided by a rosette configuration with open boundary conditions. Specifically, we consider a symmetric rosette configuration in which an isolated set of cells meet at a single point. It was shown in \cite{Spencer2017} that if all the cells have the same energy parameters and $\Gamma=0$ [see Eq. (\ref{energy0})], then a vertex connecting four cells is unstable. Here, we show that if $\Gamma>0$ this type of configuration can be stable and there is again a transition. 

For simplicity, we consider a configuration containing four cells, all connected at the same point. The four cells are located at $\vec{r}_{1}=R\hat{x}$, $\vec{r}_{2}=R\hat{y}$, $\vec{r}_{3}=-R\hat{y}$, and $\vec{r}_{4}=-R\hat{x}$, where $R\leq\ell$. Due to symmetry, the configuration is stable if
\begin{align}
&\frac{\partial E}{\partial x}=0 ,\nonumber\\
&\frac{\partial^{2}E}{\partial x^{2}}>0 ,\nonumber\\
&\frac{\partial^{2}E}{\partial y^{2}}>0 ,
\label{stabilitycond1}
\end{align}
where $x$ and $y$ are the coordinates of cell $1$, without loss of generality. Equation (\ref{stabilitycond1}) is equivalent to
\begin{align}
&\left(\left\langle \tilde{A}\right\rangle-\tilde{A}_{0}\right)\frac{\partial\left\langle \tilde{A}\right\rangle}{\partial \tilde{x}}+\tilde{K}\left(\left\langle \tilde{P}\right\rangle-\tilde{P}_{0}\right)\frac{\partial\left\langle \tilde{P}\right\rangle}{\partial \tilde{x}}=0 ,\nonumber\\
&\left(\left\langle \tilde{A}\right\rangle-\tilde{A}_{0}\right)\frac{\partial^{2}\left\langle \tilde{A}\right\rangle}{\partial \tilde{z}^{2}}+\tilde{K}\left(\left\langle \tilde{P}\right\rangle-\tilde{P}_{0}\right)\frac{\partial^{2}\left\langle \tilde{P}\right\rangle}{\partial \tilde{z}^{2}}+\nonumber\\
&+\left(\frac{\partial\left\langle \tilde{A}\right\rangle}{\partial \tilde{z}}\right)^{2}+\tilde{K}\left(\frac{\partial\left\langle \tilde{P}\right\rangle}{\partial \tilde{z}}\right)^{2}>0 ,
\end{align}
where $z$ in the second equation stands for either $x$ or $y$.

Explicitly, we find that the mean average and circumference are
\begin{align}
&\left\langle \tilde{A}\right\rangle=\frac{\tilde{R}\left(\tilde{R}+\sqrt{2-\tilde{R}^{2}}\right)+\cos^{-1}\left(-\tilde{R}\sqrt{2-\tilde{R}^{2}}\right)}{2} ,\nonumber\\
&\left\langle\tilde{P}\right\rangle=\sqrt{2}\left(\tilde{R}+\sqrt{2-\tilde{R}^{2}}\right)+\cos^{-1}\left(-\tilde{R}\sqrt{2-\tilde{R}^{2}}\right) .
\end{align}
The first derivatives with respect to $x$ are
\begin{align}
&\frac{\partial\left\langle \tilde{A}\right\rangle}{\partial\tilde{x}}=\frac{\tilde{R}+\sqrt{2-\tilde{R}^{2}}}{4} ,\nonumber\\
&\frac{\partial\left\langle \tilde{P}\right\rangle}{\partial\tilde{x}}=\frac{2-\sqrt{2}}{4}+\frac{\sqrt{2}-\tilde{R}}{2^{3/2}\sqrt{2-\tilde{R}^{2}}} .
\end{align}
The second derivatives with respect to $y$ are
\begin{align}
&\frac{\partial^{2}\left\langle \tilde{A}\right\rangle}{\partial\tilde{y}^{2}}=\frac{1-\tilde{R}^{2}}{4\tilde{R}\sqrt{2-\tilde{R}^{2}}} ,\nonumber\\
&\frac{\partial^{2}\left\langle \tilde{P}\right\rangle}{\partial\tilde{y}^{2}}=\frac{4-3\sqrt{2}}{8\tilde{R}}+\frac{4-4\sqrt{2}\tilde{R}+\sqrt{2}\tilde{R}^{3}}{8\tilde{R}\left(2-\tilde{R}^{2}\right)^{3/2}} .
\end{align}
The second derivatives with respect to $x$ are
\begin{align}
&\frac{\partial^{2}\left\langle \tilde{A}\right\rangle}{\partial\tilde{x}^{2}}=\frac{1-\tilde{R}^{2}}{4\tilde{R}\sqrt{2-\tilde{R}^{2}}} ,\nonumber\\
&\frac{\partial^{2}\left\langle \tilde{P}\right\rangle}{\partial\tilde{x}^{2}}=\frac{1}{8\tilde{R}^{2}\left(2-\tilde{R}\right)^{3/2}\left(1+\tilde{R}\sqrt{2-\tilde{R}^{2}}\right)^{2}}\nonumber\\
&\left\{\left(1+\tilde{R}\sqrt{2-\tilde{R^{2}}}\right)\left[2\tilde{R}\left(13-14\tilde{R}^{2}+3\tilde{R}^{4}\right)+\right.\right.\nonumber\\
&\left.\left.+\sqrt{2}\left(1-\tilde{R}\right)\left(23+11\tilde{R}-27\tilde{R}^{2}-17\tilde{R}^{3}+9\tilde{R}^{4}+\right.\right.\right.\nonumber\\
&\left.\left.\left.+7\tilde{R}^{5}-\tilde{R}^{6}-\tilde{R}^{7}\right)\right]-\left(1-\tilde{R}^{2}\right)^{2}\left[2\tilde{R}\left(3-2\tilde{R}^{3}\right)+\right.\right.\nonumber\\
&\left.\left.+\sqrt{2}\left(1-\tilde{R}\right)\left(11+7\tilde{R}-8\tilde{R}^{2}-6\tilde{R}^{3}+\tilde{R}^{4}+\tilde{R}^{5}\right)\right]\right\}
\end{align}

\begin{figure}
\subfigure[$\tilde{K}=1$]{\includegraphics[width=0.4\columnwidth]{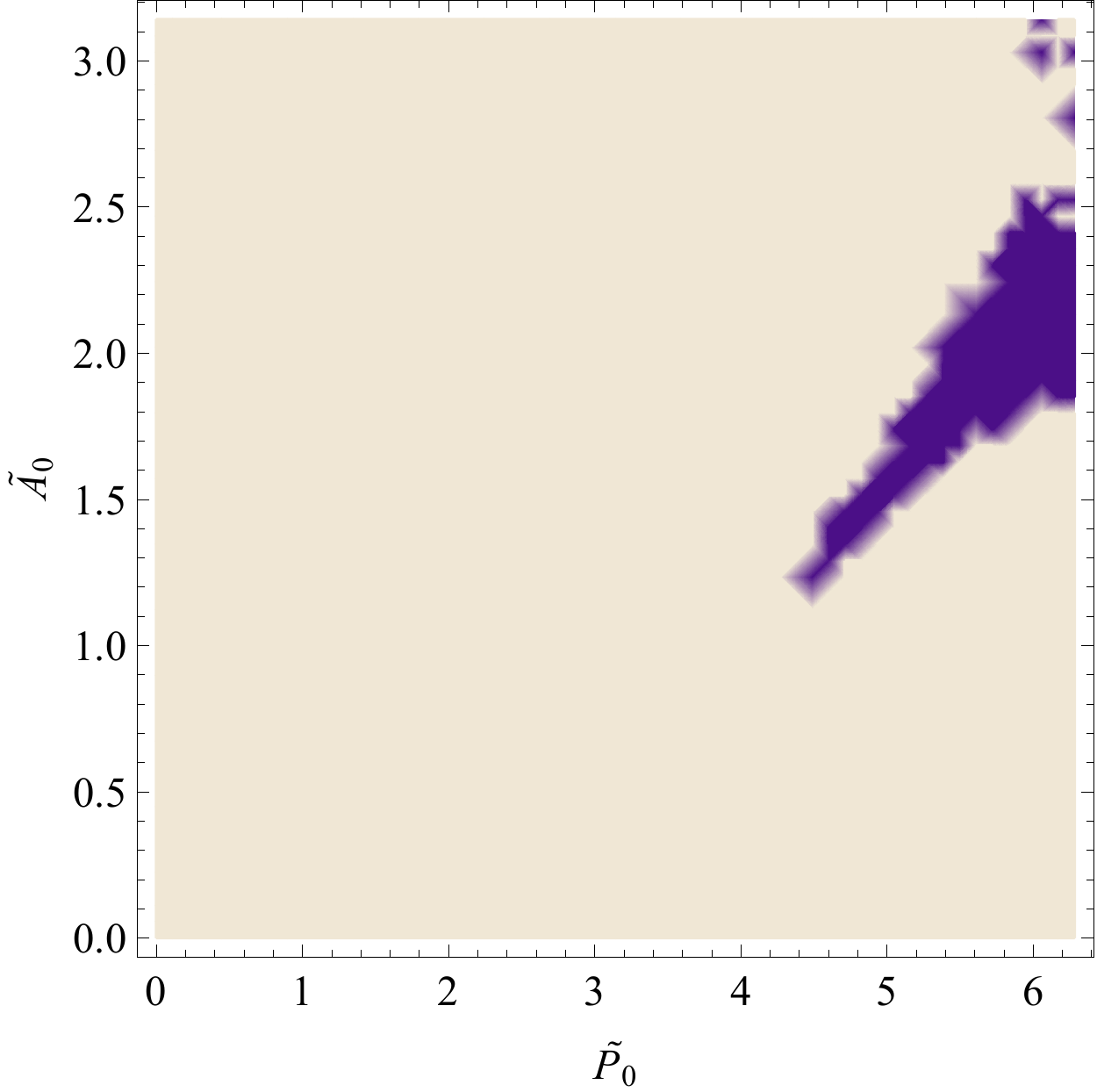}}
\subfigure[$\tilde{K}=10$]{\includegraphics[width=0.4\columnwidth]{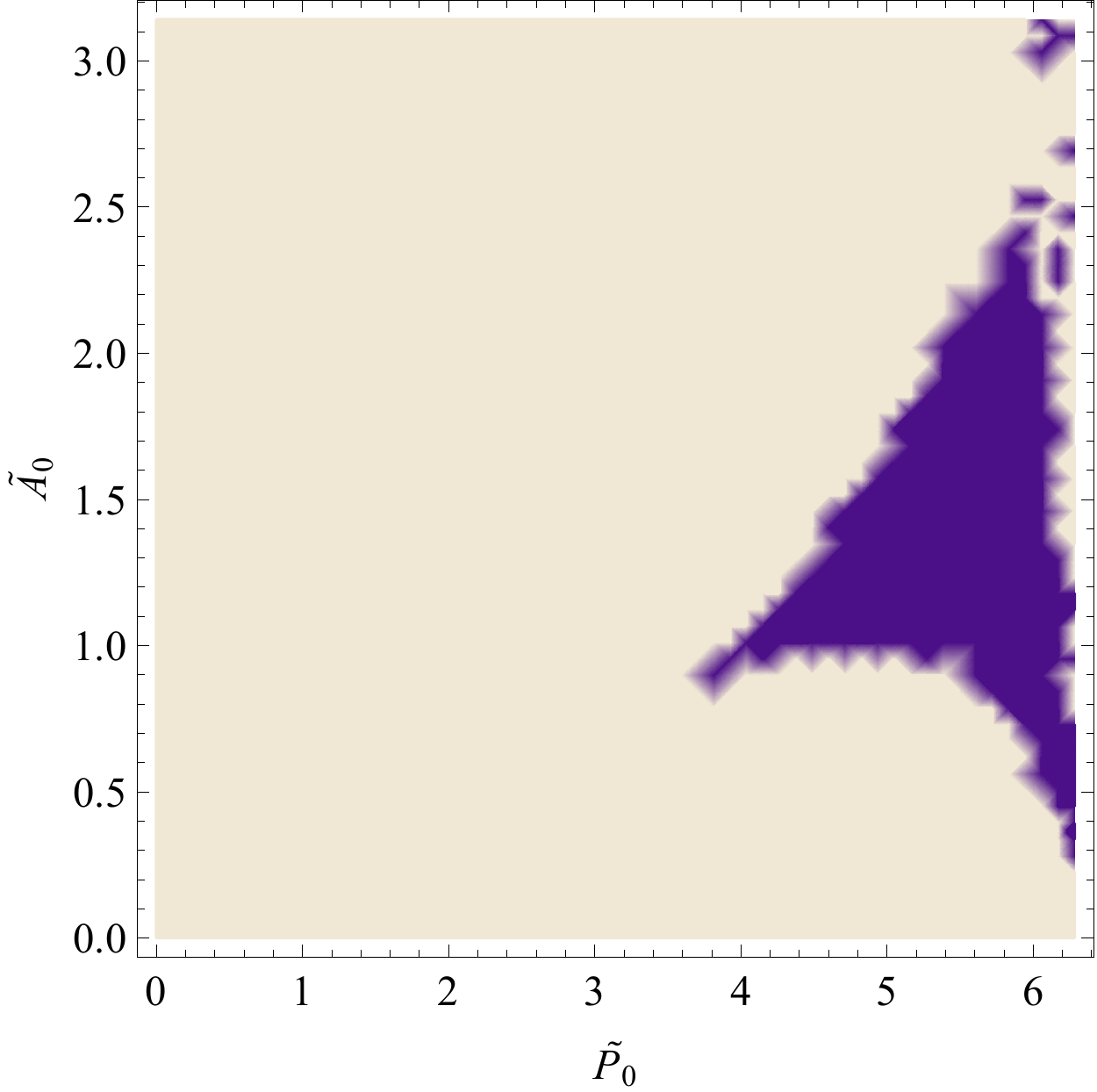}}
\caption{A phase diagram in the $\tilde{A}_{0}-\tilde{P}_{0}$ plane showing the stability of the four-cell vertex configuration, for two values of $\tilde{K}$. The configuration is stable in the blue regions, and unstable in the white regions.}
\label{4cell_stable}
\end{figure}

We solve $\frac{\partial E}{\partial x}=0$ numerically for $\tilde{R}$ and set the result in the two inequalities for the second derivatives. The configuration is stable if $0<\tilde{R}<1$ and the two second derivatives are positive. Figure \ref{4cell_stable} shows in blue the region in the $\tilde{A}_{0}-\tilde{P}_{0}$ plane for which the configuration is stable, for two values of $\tilde{K}$.

\subsection{Differences between the honeycomb lattice and the rosette pattern}

There are several qualitative differences of the rosette pattern from the honeycomb lattice. First is the dependence of the transition line on the value of $K_{A}/\left(K_{P}A_{0}\right)$. Second, increasing the value of $A_{0}$ causes the rosette configuration to eventually become unstable, while it causes the honeycomb lattice to be more stable.

\section{Phase Diagram}
\label{sec_cluster}

\begin{figure}
\centering
\subfigure[Gas ({\bf{G}}) ]{\includegraphics[width=0.32\columnwidth]{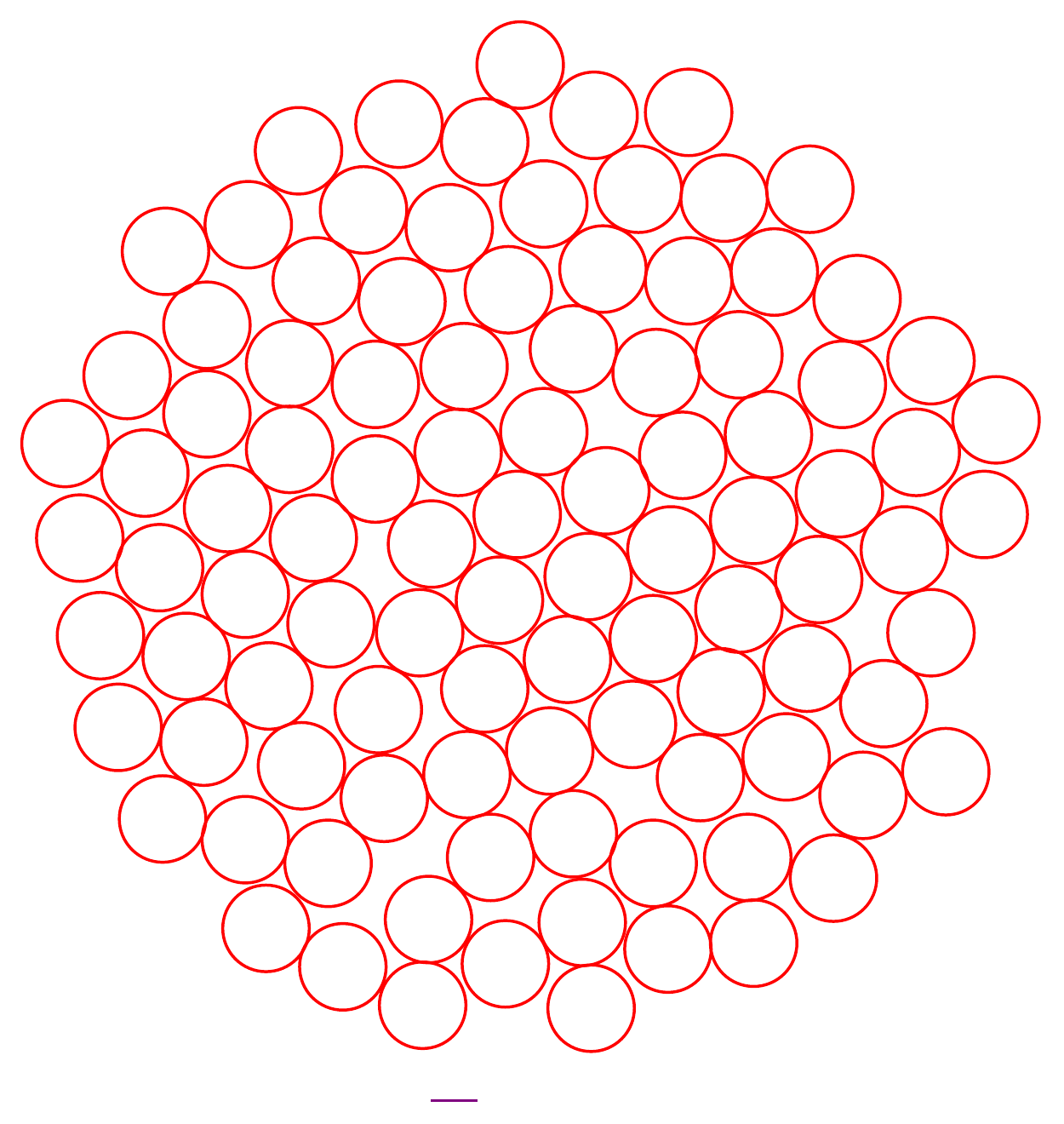}}
\subfigure[Cluster\,({\bf{Cl}})]{\includegraphics[width=0.32\columnwidth]{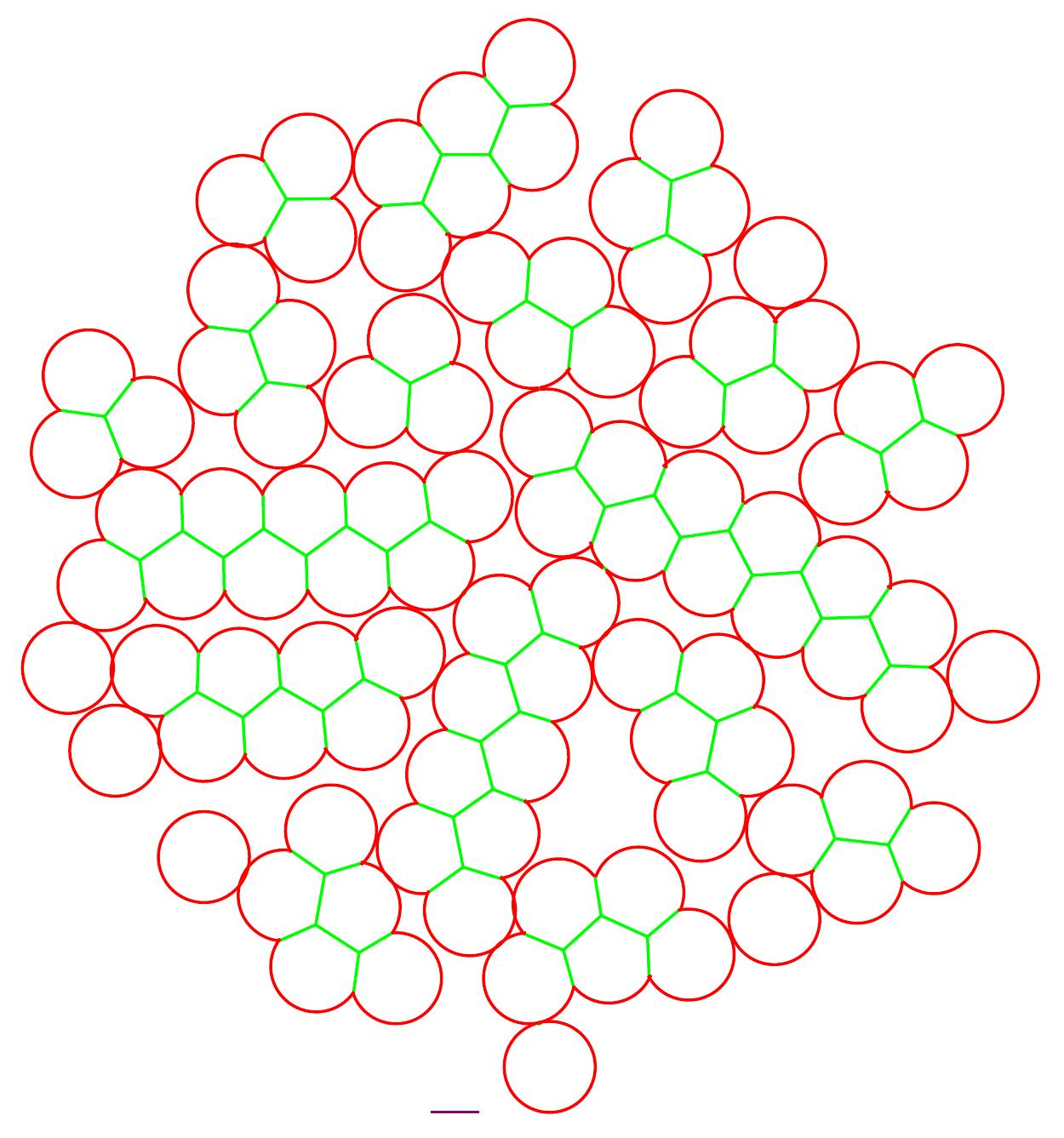}}
\subfigure[Hexagonal ({\bf{H}})]{\includegraphics[width=0.32\columnwidth]{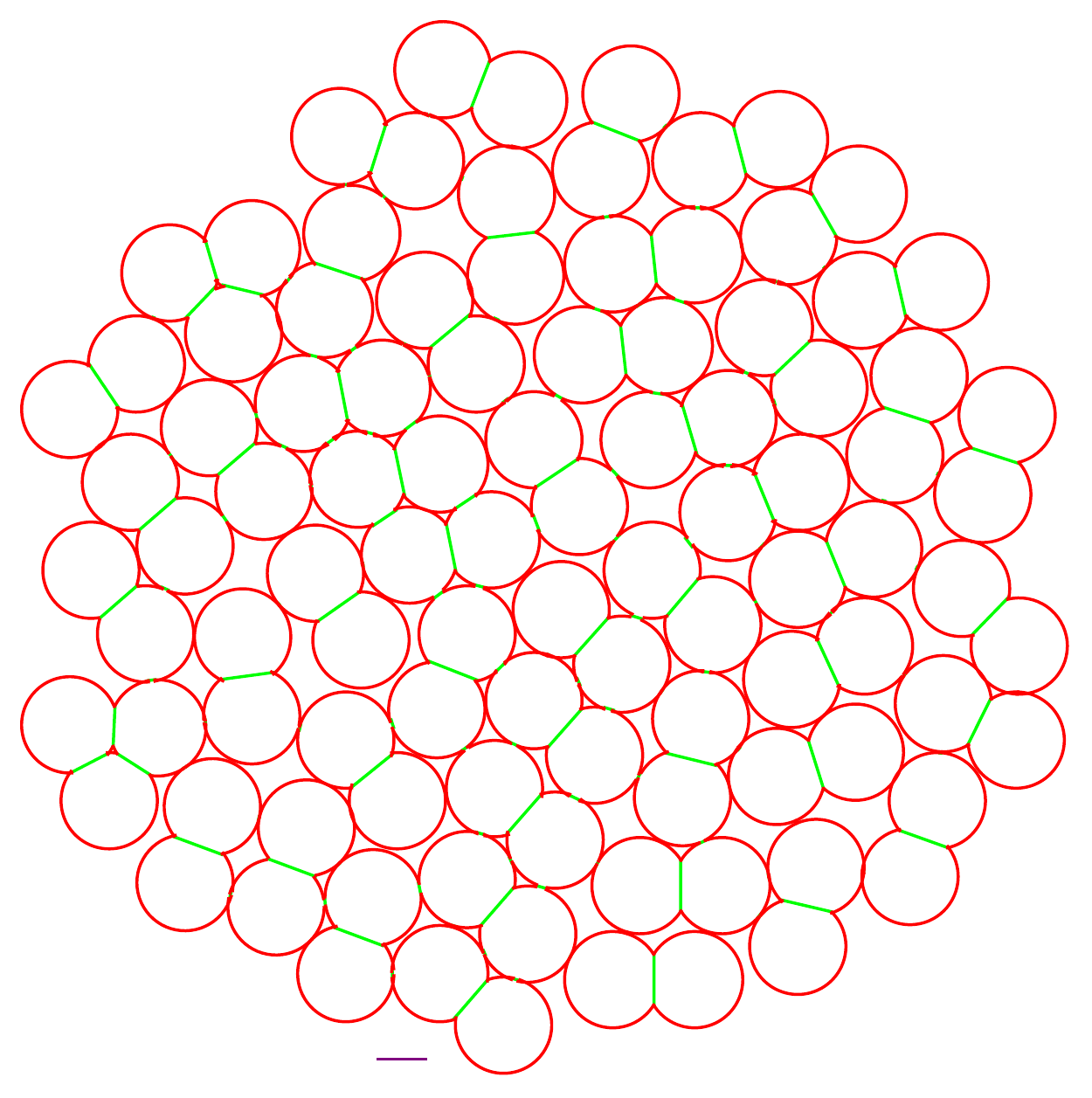}}\\
\subfigure[Non-Confluent ({\bf{NC}})]{\includegraphics[width=0.35\columnwidth]{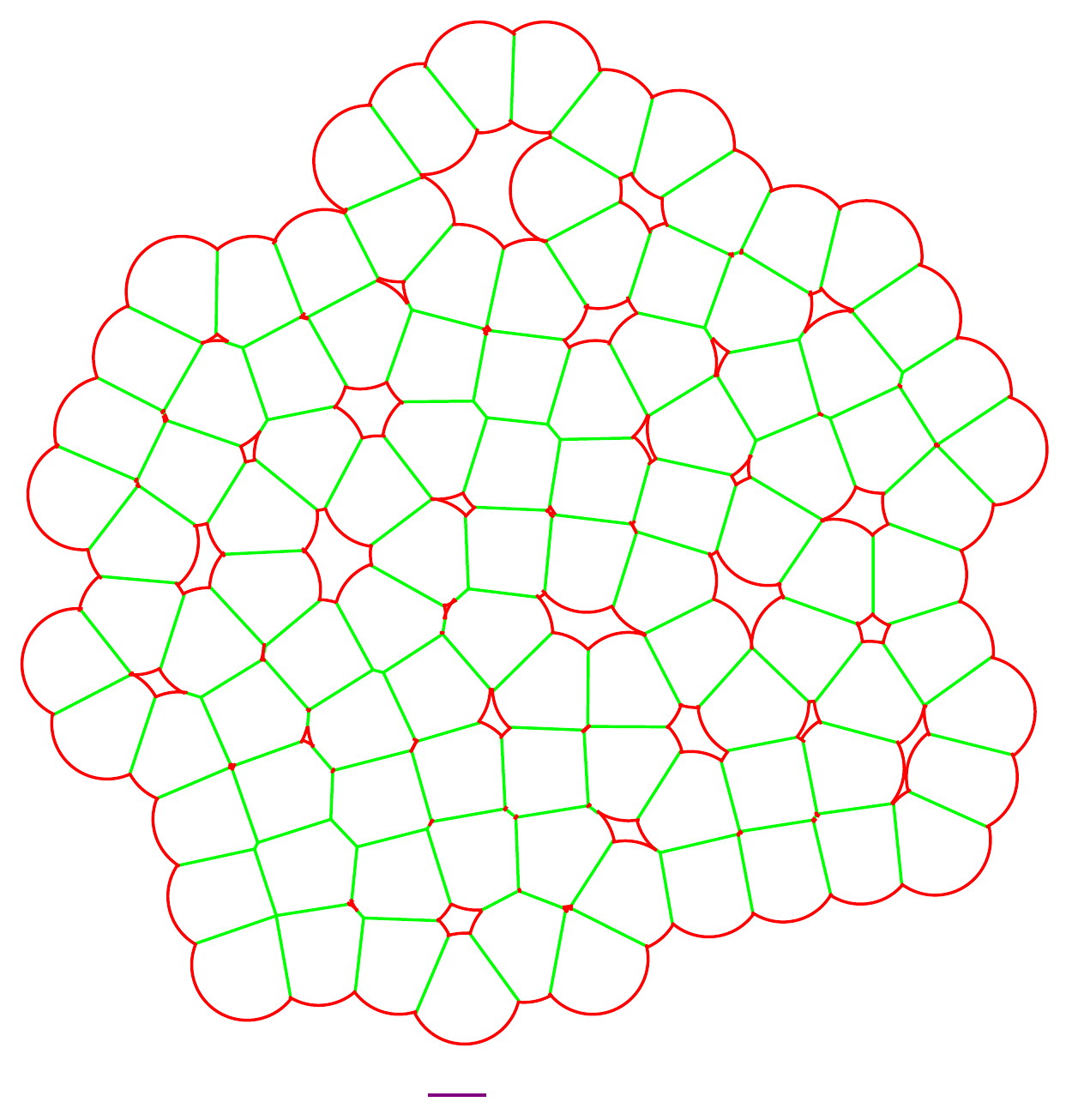}}
\subfigure[Minimal ({\bf{M}})]{\includegraphics[width=0.31\columnwidth]{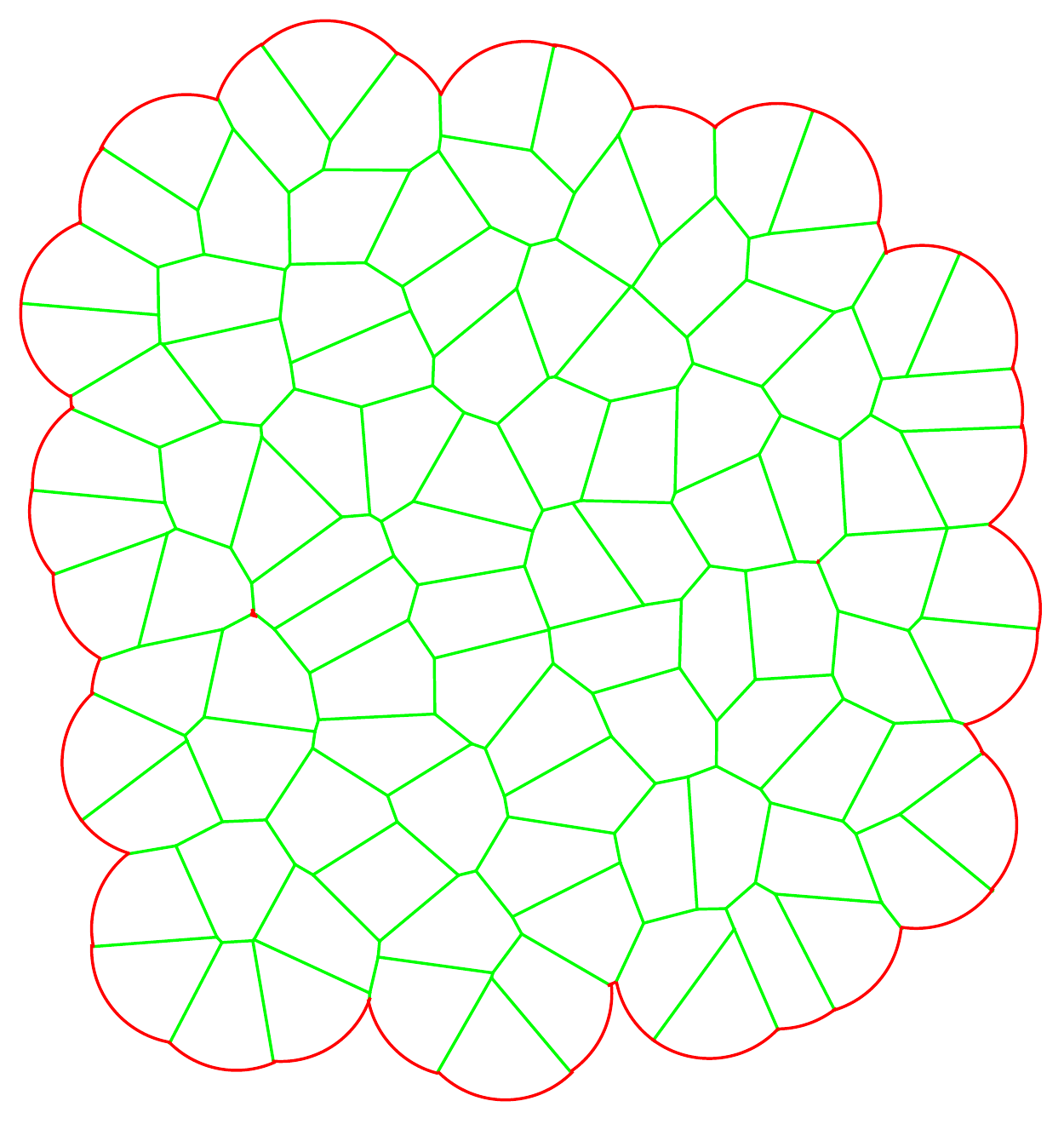}}
\subfigure[Confluent ({\bf{Co}})]{\includegraphics[width=0.31\columnwidth]{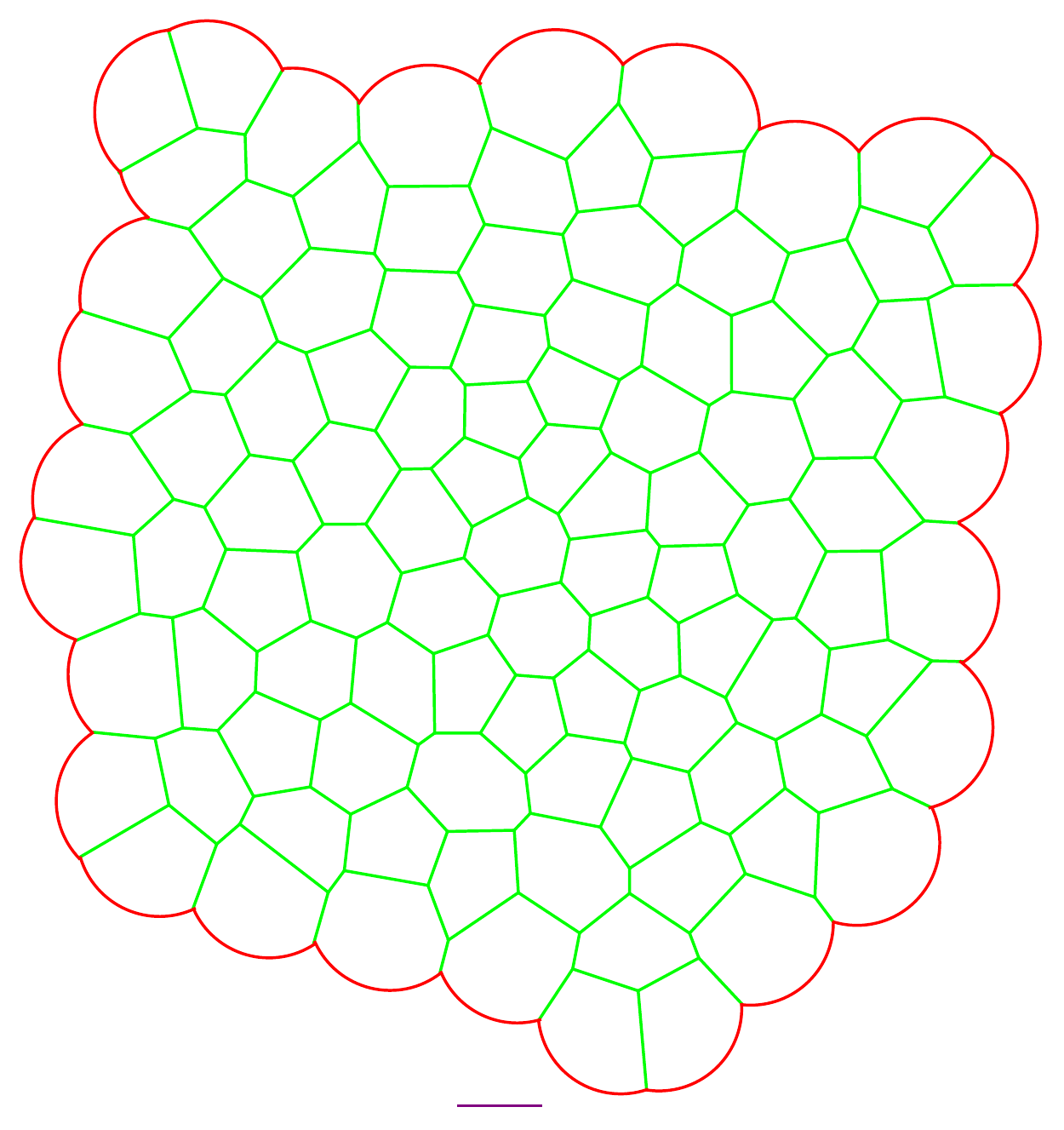}}\\
\caption{Simulations results for the state of the system after a long time exemplifying each of the phases, starting from the same initial condition with $N=100$ cells and zero temperature $T=0$. Green lines are boundaries between adjacent cells, and red lines are the outer boundaries of the cells. The purple line below each configuration is of length $\ell=1$ to see the different scales. The parameters used are $\tilde{K}=1$, and (a) $\tilde{A}_{0}=6$ and $\tilde{P}_{0}=5$; (b) $\tilde{A}_{0}=4.5$ and $\tilde{P}_{0}=4.3$; (c) $\tilde{A}_{0}=3.1$ and $\tilde{P}_{0}=6.1$; (d) $\tilde{A}_{0}=1$ and $\tilde{P}_{0}=8$; (e) $\tilde{A}_{0}=1$ and $\tilde{P}_{0}=4$; (f) $\tilde{A}_{0}=4$ and $\tilde{P}_{0}=2.5$.}
\label{fig_clusters}
\end{figure}

\begin{figure}[t]
\centering
\includegraphics[width=0.9\columnwidth]{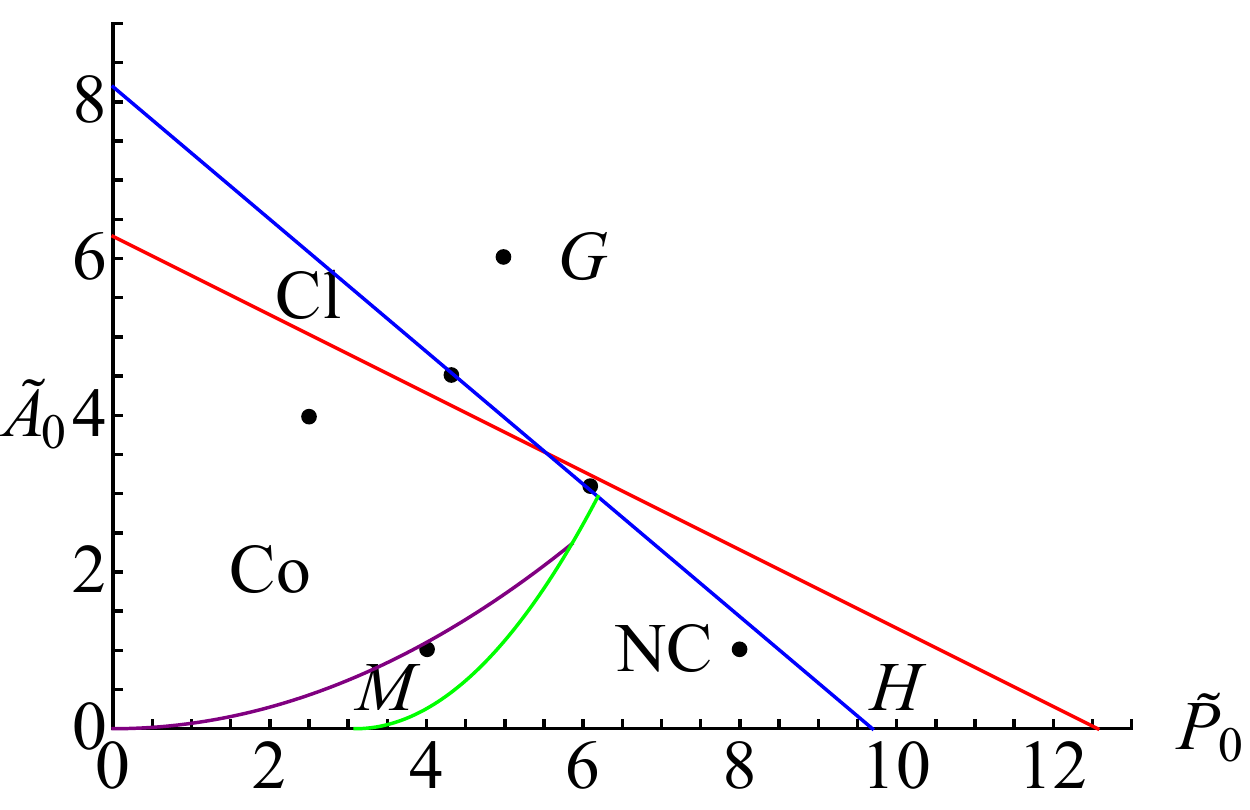}
\caption{The phase diagram of fixed open boundary conditions for $\tilde{K}=1$. It is qualitatively the same for all values of $\tilde{K}$. The letters correspond to the different phases Gas ({\bf{G}}), Cluster ({\bf{Cl}}), Hexagonal ({\bf{H}}), Non-Confluent ({\bf{NC}}), Minimal ({\bf{M}}), and Confluent ({\bf{Co}}). The lines are boundaries between the phases: Eq. (\ref{2cellint}) in red, Eq. (\ref{3cellint}) in blue, $\tilde{P}_{0}=q_{5}\sqrt{\tilde{A}_{0}}$ in purple, and Eq. (\ref{pastapp}) in green. The dot in each phase corresponds to the parameters of the configurations shown in Fig. \ref{fig_clusters}.}
\label{fig_phases}
\end{figure}

In our model, with open boundary conditions the cells do not necessarily comprise one connected cluster. We next investigate the different phases exhibited for different $\tilde{P}_0$ and $\tilde{A}_0$ by these connected or unconnected clusters. To begin this study, sweeping through the $\tilde{P}_0$ and $\tilde{A}_0$ plane, with $\tilde{K}=1$, we ran simulations (for fixed $\ell=1$) at zero temperature starting with 100 cells (i.e. reference points) placed at random with a density much larger than $1/\ell^2$. We then ran the dynamics until the system reached equilibrium. We found that in different regions of the $\tilde{P}_0$, $\tilde{A}_0$ plane, the system adopted distinctive configurations, which we identify as different phases. This set of typical configurations is shown in Fig. \ref{fig_clusters}. A phase diagram detailing the locations of the different configurations is shown in Fig. \ref{fig_phases}. The properties of the configurations obtained from the numerical simulations agree perfectly with the analytically derived phase diagram shown in Fig. \ref{fig_phases}. We also ran simulations for different values of $\tilde{K}$ and found qualitatively similar results.

We can understand the structure of the phase diagram analytically by considering two-cell interactions, three-cell interactions, and geometrical constraints.

\subsection{Two-cell and three-cell interactions}
As a rough approximation for two-cell contacts, we consider two cells located at $\vec{r}_{\pm}=\pm\left(\ell-\epsilon\right)\hat{x}$, and for three-cell contacts, we consider three cells located on the vertices of an equilateral triangle with side length $\sqrt{3}(\ell-\epsilon)$, such that when $\epsilon=0$ the three cells touch at one point. Note that in both cases $\epsilon\geq0$. The three cells are located at
\begin{align}
&\vec{r}_{0}=\left(\ell-\epsilon\right)\hat{y} ,\nonumber\\
&\vec{r}_{\pm}=\left(\ell-\epsilon\right)\left(-\frac{\hat{y}}{2}\pm\frac{\sqrt{3}\hat{x}}{2}\right) .
\end{align}
The contacts are attractive if
\begin{align}
\left.\frac{\partial E}{\partial\epsilon}\right|_{\epsilon=0}<0 .\label{dee}
\end{align}

In the two cell configuration, the area and circumference of each cell are
\begin{align}
&A=\left(\ell-\epsilon\right)\sqrt{\ell^{2}-\left(\ell-\epsilon\right)^{2}}+\frac{\ell^{2}\theta}{2} ,\nonumber\\
&P=2\sqrt{\ell^{2}-\left(\ell-\epsilon\right)^{2}}+\ell\theta .
\end{align}
The force acting on the cells is
\begin{align}
\tilde{f}=-2\sqrt{\tilde{\epsilon}\left(2-\tilde{\epsilon}\right)}\left[\left(\tilde{A}-\tilde{A}_{0}\right)+\tilde{K}\frac{\left(\tilde{P}-\tilde{P}_{0}\right)}{2-\tilde{\epsilon}}\right] .
\end{align}
At $\epsilon\approx0$, the force may be approximated by
\begin{align}
&\tilde{f}\left(\tilde{\epsilon}\approx0\right)\approx\sqrt{2\epsilon}\left[2\tilde{A}_{0}-2\pi+\tilde{K}\left(\tilde{P}_{0}-2\pi\right)\right] .\nonumber\\
\end{align}
Therefore, the two-cell interaction is attractive if
\begin{align}
2\left(\pi-\tilde{A}_{0}\right)+\tilde{K}\left(2\pi-\tilde{P}_{0}\right)>0 .\label{2cellint}
\end{align}

In the three cell configuration, the area and circumference of the cells are
\begin{align}
&A=\frac{\sqrt{3}\left(\ell-\epsilon\right)\left(\ell-\epsilon+\sqrt{4\ell^{2}-3\left(\ell-\epsilon\right)^{2}}\right)}{4}+\ell^{2}\theta_{3} ,\nonumber\\
&P=\ell-\epsilon+\sqrt{4\ell^{2}-3\left(\ell-\epsilon\right)^{2}}+\ell\theta_{3} ,
\end{align}
with
\begin{align}
&\theta_{3}=2\pi-\cos^{-1}\left[\frac{3}{4}\left(1-\frac{\epsilon}{\ell}\right)^{2}-\frac{1}{2}-\right.\nonumber\\
&\left.-\frac{3}{4}\left(1-\frac{\epsilon}{\ell}\right)\sqrt{4-3\left(1-\frac{\epsilon}{\ell}\right)^{2}}\right] .\label{arp3}
\end{align}
Using Eqs. (\ref{dee}) and (\ref{arp3}) we find that the three-cell interaction is attractive if
\begin{align}
&\left(\frac{3\sqrt{3}+4\pi}{6}-\tilde{A}_{0}\right)+\nonumber\\
&+\frac{2}{3}\left(3-\sqrt{3}\right)\tilde{K}\left(\frac{2\left(3+2\pi\right)}{3}-\tilde{P}_{0}\right)>0 .\label{3cellint}
\end{align}

Each of the two interactions can be independently either attractive or repulsive. If both of them are repulsive, all the cells repel each other and the system is in the Gas phase, marked as {\bf{G}} in Fig. \ref{fig_phases} and exemplified in Fig. \ref{fig_clusters}a. If the three-cell interaction is attractive and the two-cell interaction is repulsive, then at low temperatures clusters in which all the cells are connected by three-cell contacts are stable. These clusters repel each other, since at low temperatures the only contact between them is via two-cell contacts. There might also be clusters with fingers that do not touch each other. We call this the Cluster phase, marked as {\bf{Cl}} in   Fig. \ref{fig_phases} and a representative sample of which is displayed in Fig. \ref{fig_clusters}b. If the three-cell interaction is repulsive and the two-cell interaction is attractive, the cells are connected in a single cluster. Whenever three or more cells meet at a single point it is always at a distance $\ell$ from each other. The reason is that if they do get closer, the interactions between them turns from three two-cell interactions, to one three-cell interaction which is repulsive. Due to this constraint, they tend to be arranged in a hexagonal lattice at low temperature, and we thus name this phase the Hexagonal phase ({\bf{H}}), shown in Fig \ref{fig_clusters}c. If both types of interactions are attractive, the cells aggregate in a single cluster, which may appear in several different phases.

We note in passing that inside the Gas phase and near the Gas-Cluster transition, there are rare stable structures that include more than three cells. However, as these types of stable structures are rare, their net effect on the global parameters is negligible, so we do not consider this a different phase. See Appendix \ref{5cell} for an example of such a structure.

\subsection{Geometric constraints}
 
In addition to the key role of the two-cell and three-cell interactions, we identify two other critical $\tilde{K}$-independent curves in the $\tilde{A}_{0}-\tilde{P}_{0}$ space. The first one is connected to the  aforementioned transition in the periodic honeycomb lattice. A closely related transition curve was found in confluent models and it lies at $\frac{\tilde{P}_{0}}{\sqrt{\tilde{A}_{0}}}=q_{5}$, where
\begin{align}
q_{n}=\sqrt{4n\tan\left(\frac{\pi}{n}\right)}
\end{align}
is the shape parameter of a regular polygon with $n$-sides \cite{Farhadifar2007,Bi2015} and $q_{5}\approx3.8$. The fact the transition depends on $q_5$ instead of $q_6$ is due to the prevalence of pentagonal cell shapes, as seen for example in Fig. \ref{fig_clusters}f. This has been previously noted (under periodic boundary conditions) both in a Voronoi-based model \cite{Bi2016} and in a vertex model \cite{Farhadifar2007,Bi2015}.

To the right of this curve, we find the Minimal ({\bf{M}}) phase. In the Minimal phase, we have $\langle\tilde{P}\rangle=\tilde{P}_{0}$ and $\langle\tilde{A}\rangle=\tilde{A}_{0}$, and hence the total energy of the system is equal to $0$. Note also that in the Minimal phase there is an abundance of vertices connecting four or more cells. As discussed above, these rosette type configurations are stable, and this is correlated with the lack of a $T1$ transition barrier. To the left of this curve is the phase labeled as Confluent ({\bf{Co}}). Here, we find from the simulation that the cells do not satisfy this zero energy condition. This is due to the geometrical constraint determining the minimal area for a given perimeter~\cite{Staple2010} and here there are many local minima, indicative of frustration. The cells instead  aggregate into confluent clusters. We also observe that in this region
\begin{align}
\langle\tilde{P}\rangle_{in} /\sqrt{\langle\tilde{A}\rangle_{in}}=q_{5} ,
\end{align}
where $\langle\cdot\rangle_{in}$ is the average over all cells which are not on the boundary. This is shown in Fig. \ref{fig_ap}b.  It is straightforward to check that what we have labeled as the Cluster phase is always within the overall Confluent region.

\begin{figure}
\includegraphics[width=0.4\columnwidth]{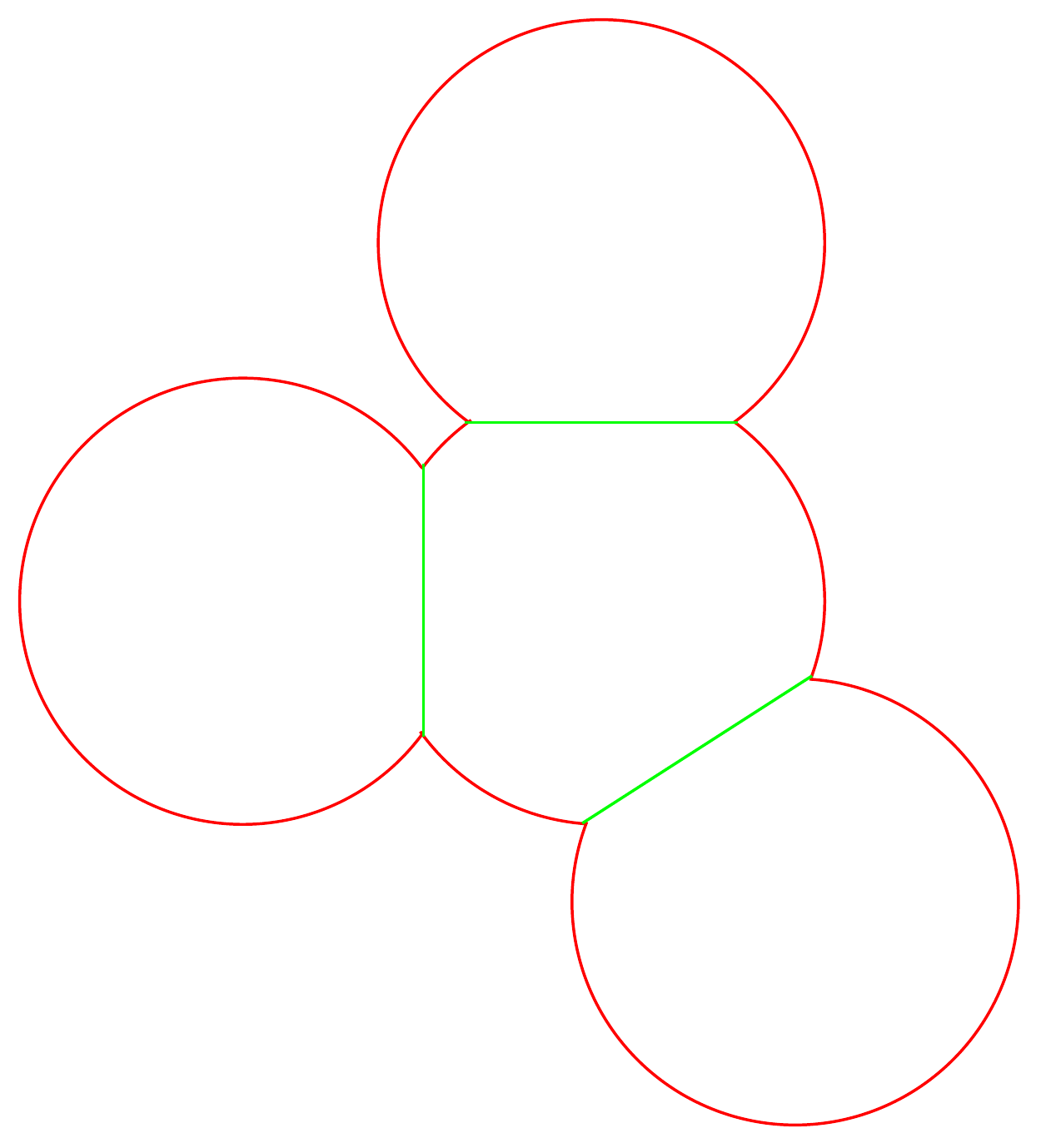}
\caption{An illustration of a cell surrounded by $3$ non-overlapping cells.}
\label{4cell_nc}
\end{figure}

In a standard vertex model, the {\bf M} phase can extend to large $\tilde{P}_{0}$. Here, however, a second phase-boundary curve occurs when due to the finite radius of the cells $\ell$, they cannot adjust their shape to reach the global energy minimum in which $P_{i}=P_{0}$ and $A_{i}=A_{0}$. We can describe this curve approximately as the locus of perimeter and area values for a cell around which we place $n$ non-overlapping other cells (i.e. cells separated by at least $2 \ell$) which do overlap with the original cell, see Fig. \ref{4cell_nc} for an illustration. 

\begin{figure}
\includegraphics[width=\columnwidth]{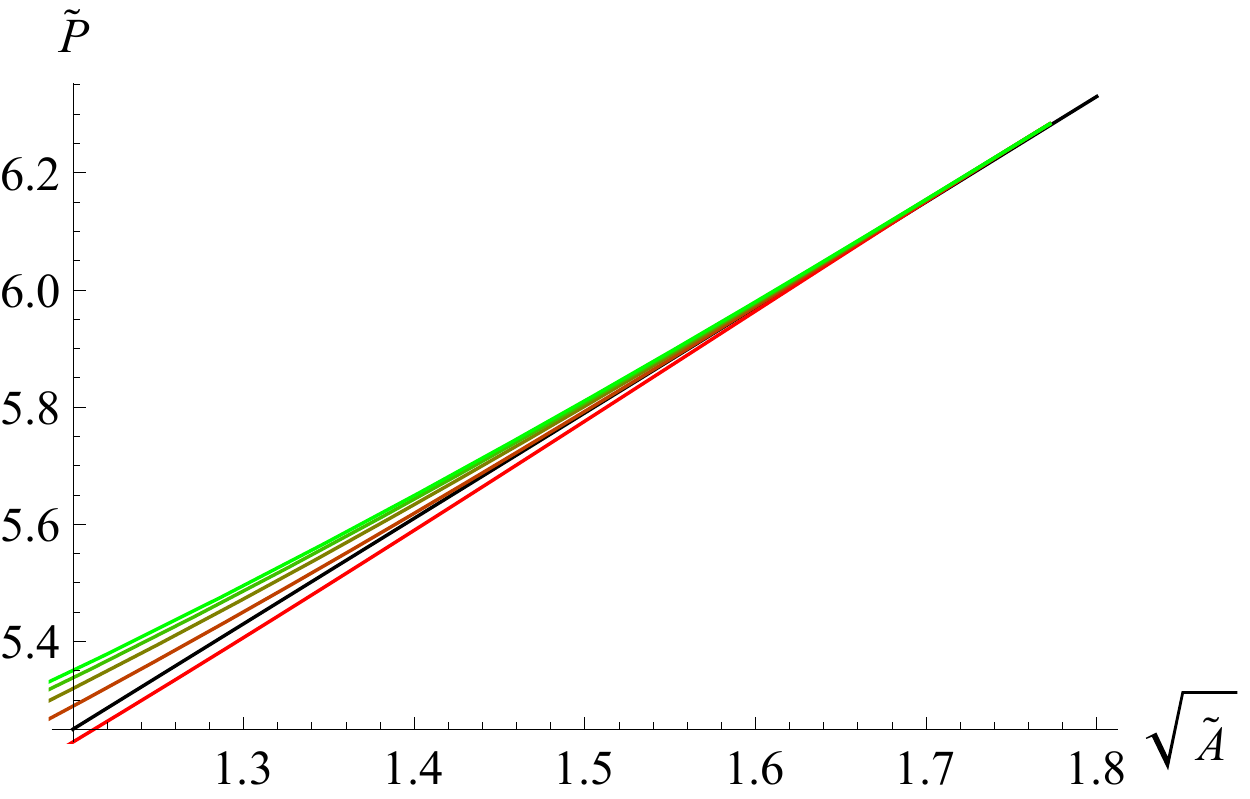}
\caption{The relation between the circumference $\tilde{P}$ and the square root of the area, $\tilde{A}$, for a cell in the non-confluent phase under the mean field approximation. The colored lines correspond to different values of $n$ from $n=2$ (red) to $n=6$ (green). The black line is $\tilde{P}=3.09+1.80\sqrt{\tilde{A}}$.}
\label{nonconfluent}
\end{figure}

\begin{figure}
\centering
\subfigure[Non-Confluent]{\includegraphics[width=0.47\columnwidth]{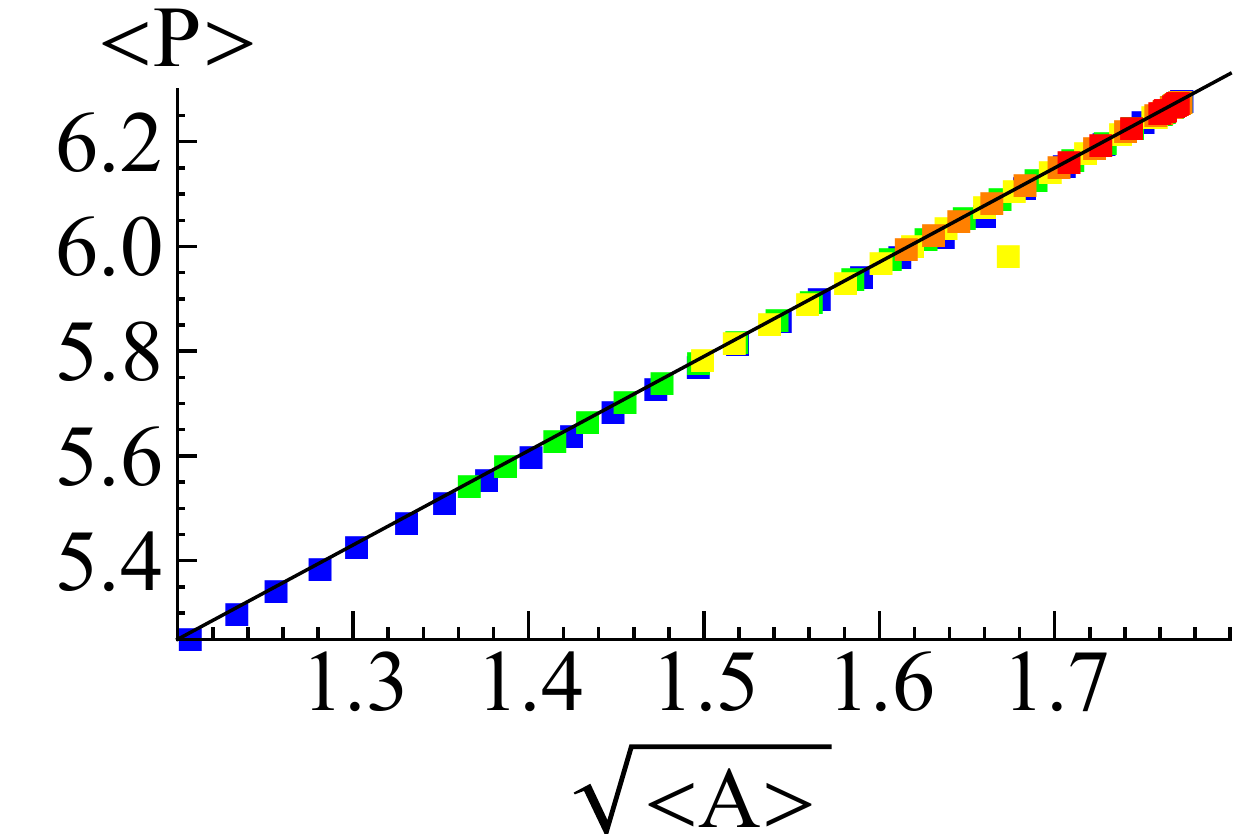}}
\subfigure[Confluent]{\includegraphics[width=0.47\columnwidth]{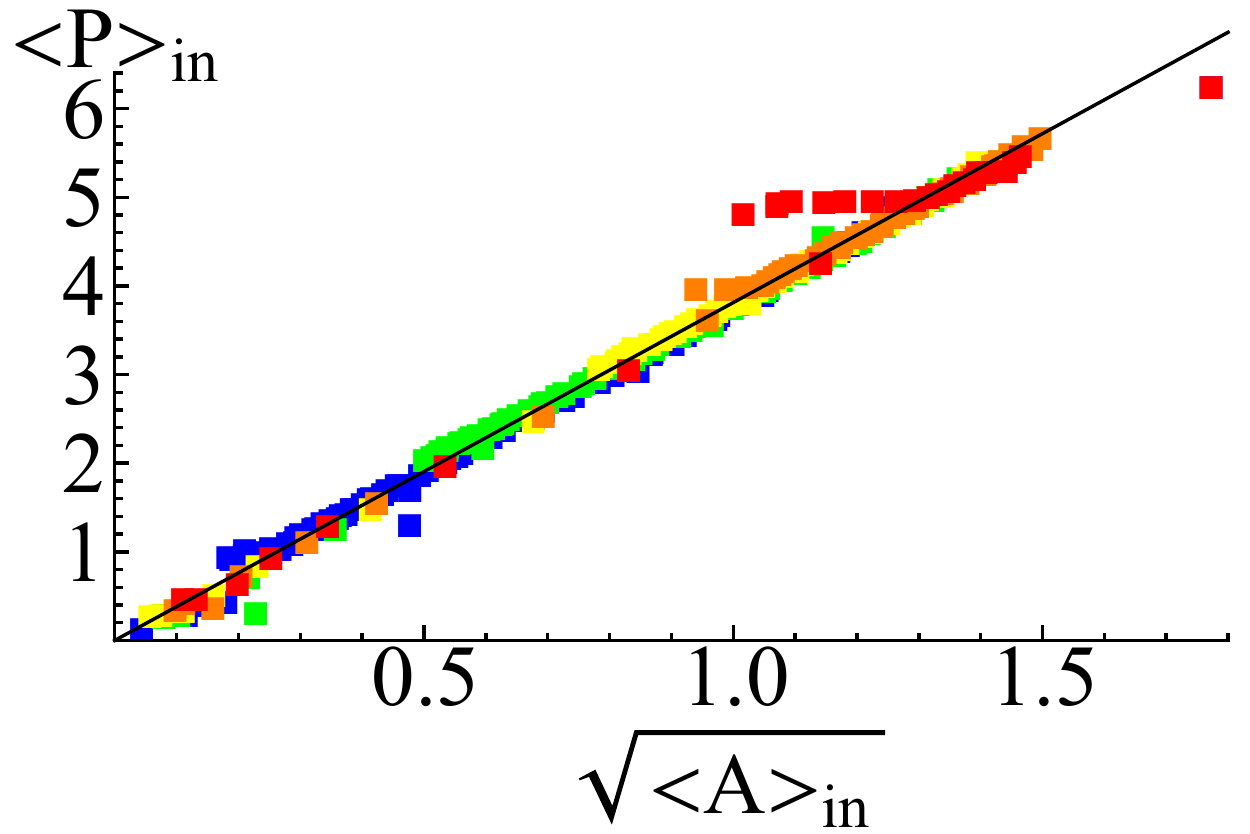}}\\
\caption{Simulation results showing the behavior of the Confluent and Non-Confluent phases. (a) A parametric plot of the mean circumference $\langle\tilde{P}\rangle$ vs. the square root of the mean area $\langle\tilde{A}\rangle^{1/2}$, for $\tilde{K}=1$ and various values of $\tilde{A}_{0}$ and $\tilde{P}_{0}$. Different colors correspond to different values of $\tilde{P}_{0}$ according to $\tilde{P}_{0}=6$ (blue), $7$ (green), $8$ (yellow), $9$ (orange), and $10$ (red). The black line is $\left\langle\tilde{P}\right\rangle=3.09+1.80\left\langle\tilde{A}\right\rangle^{1/2}$. (b) A parametric plot of the mean circumference of inner cells $\langle\tilde{P}\rangle_{in}$ vs. the square root of the mean area of the inner cells $\langle\tilde{A}\rangle_{in}^{1/2}$, for $\tilde{K}=1$ and various values of $\tilde{A}_{0}$ and $\tilde{P}_{0}$. Different colors correspond to different values of $\tilde{P}_{0}$ according to $\tilde{P}_{0}=1$ (blue), $2$ (green), $3$ (yellow), $4$ (orange), and $5$ (red). The black line is $\langle\tilde{P}\rangle_{in}=q_{5}\langle\tilde{A}\rangle_{in}^{1/2}$.}
\label{fig_ap}
\end{figure}

The area and circumference of the central cell are
\begin{align}
&\tilde{A}=\pi-n\left[\frac{\cos^{-1}\left(\frac{\tilde{R}^{2}}{2}-1\right)}{2}-\frac{\tilde{R}\sqrt{4-\tilde{R}^{2}}}{4}\right] ,\nonumber\\
&\tilde{P}=2\pi-n\left[\cos^{-1}\left(\frac{\tilde{R}^{2}}{2}-1\right)-\sqrt{4-\tilde{R}^{2}}\right] .
\end{align}
Figure \ref{nonconfluent} is a parametric plot of $\tilde{P}$ vs. $\sqrt{\tilde{A}}$ as $\tilde{R}$ is varied from $0$ to $2$, for values of $n$ ranging from $2$ to $6$. In the region of interest, we find that this relation is very well approximated by
\begin{align}
\tilde{P}\approx3.09+1.80\sqrt{\tilde{A}} .\label{pastapp}
\end{align}

Beyond this second curve lies the region we refer to as the Non-Confluent ({\bf NC}) phase, shown in Fig. \ref{fig_clusters}d. Inside the region, we find that $\left\langle\tilde{P}\right\rangle$ and $\left\langle\tilde{A}\right\rangle$ satisfy Eq. (\ref{nonconfluent}). This is shown in Fig. \ref{fig_ap}a.

\subsection{Order parameters}

The transitions between the different phases can also be quantified by looking at the discontinuities of the derivatives of various order parameters with respect to the control parameters, $\tilde{A}_{0}$, $\tilde{P}_{0}$ and $\tilde{K}$. The Minimal-Confluent and Minimal-Non Confluent transitions involve a discontinuity in the derivative of the energy, since in the Minimal phase the energy is equal to zero, while in the other two phases it depends nontrivially on the control parameters. 

\begin{figure}
\includegraphics[width=0.9\columnwidth]{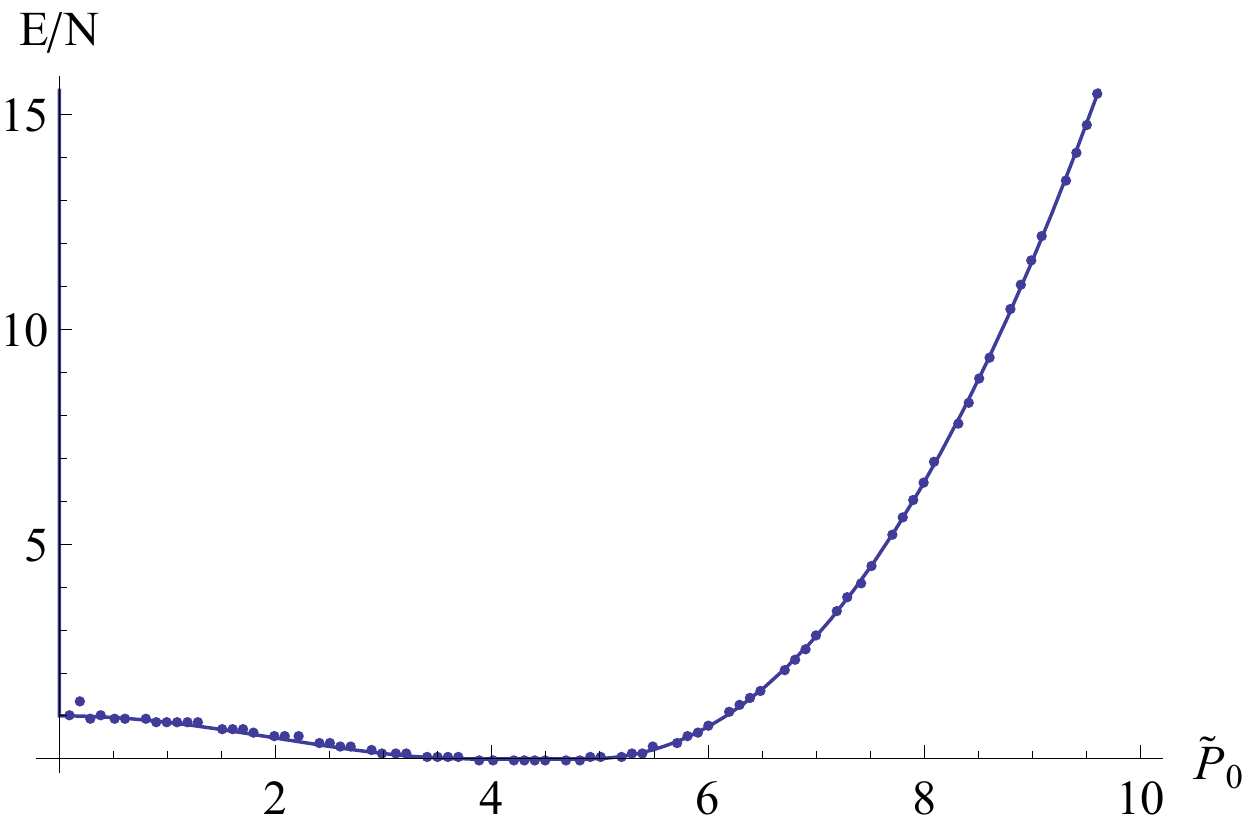}
\caption{The mean energy per cell vs. $\tilde{P}_{0}$ for $\tilde{K}=\tilde{A}_{0}=1$. The symbols are the numerical results, and the continuous line is the analytical expression obtained from Eqs. (\ref{energy1}), (\ref{mean_p}) and (\ref{mean_a}).}
\label{fig_energy}
\end{figure}

In order to approximate  the energy in the Confluent and Non-Confluent phases, we note that we numerically found that in those phases the variances of the area and circumference are very small, and we thus assume that $\langle\tilde{P}^{2}\rangle\approx\langle\tilde{P}\rangle^{2}$ and $\langle\tilde{A}^{2}\rangle\approx\langle\tilde{A}\rangle^{2}$. Since we know the relation between the mean circumference and area, we may write the total energy in terms of $\langle\tilde{A}\rangle$ and minimize it. In the Confluent phase, the area-perimeter relationship is taken to be that of a perfect pentagon, while in the Non-Confluent phase we use Eq. (\ref{pastapp}). We find that the mean circumference is the solution of the cubic equation
\begin{align}
&\frac{2}{c^{4}_{2}}\left(\langle\tilde{P}\rangle-c_{1}\right)^{3}+\left(\tilde{K}-\frac{2\tilde{A}_{0}}{c^{2}_{2}}\right)\left(\langle\tilde{P}\rangle-c_{1}\right)-\nonumber\\
&-\tilde{K}\left(\tilde{P}_{0}-c_{1}\right)=0 ,\label{mean_p}
\end{align} 
and the mean area is the solution of the equation
\begin{align}
&2\langle\tilde{A}\rangle-2\tilde{A}_{0}+\tilde{K}c^{2}_{2}-\frac{c_{2}\tilde{K}\left(\tilde{P}_{0}-c_{1}\right)}{\sqrt{\langle\tilde{A}\rangle}}=0 ,\label{mean_a}
\end{align}
where in the Confluent phase $c_{1}=0$ and $c_{2}=q_{5}$, while in the Non-Confluent phase $c_{1}=3.09$ and $c_{2}=1.80$.
As shown in Fig. \ref{fig_energy}, we find a nearly perfect agreement with the numerical results in the Non-Confluent phase, and a lower bound on the energy in the Confluent phase. The latter is to be expected, as the approximate treatment as  perfect pentagons invariably breaks down near the boundaries, where the cells have higher energies.

\begin{figure}
\includegraphics[width=0.9\columnwidth]{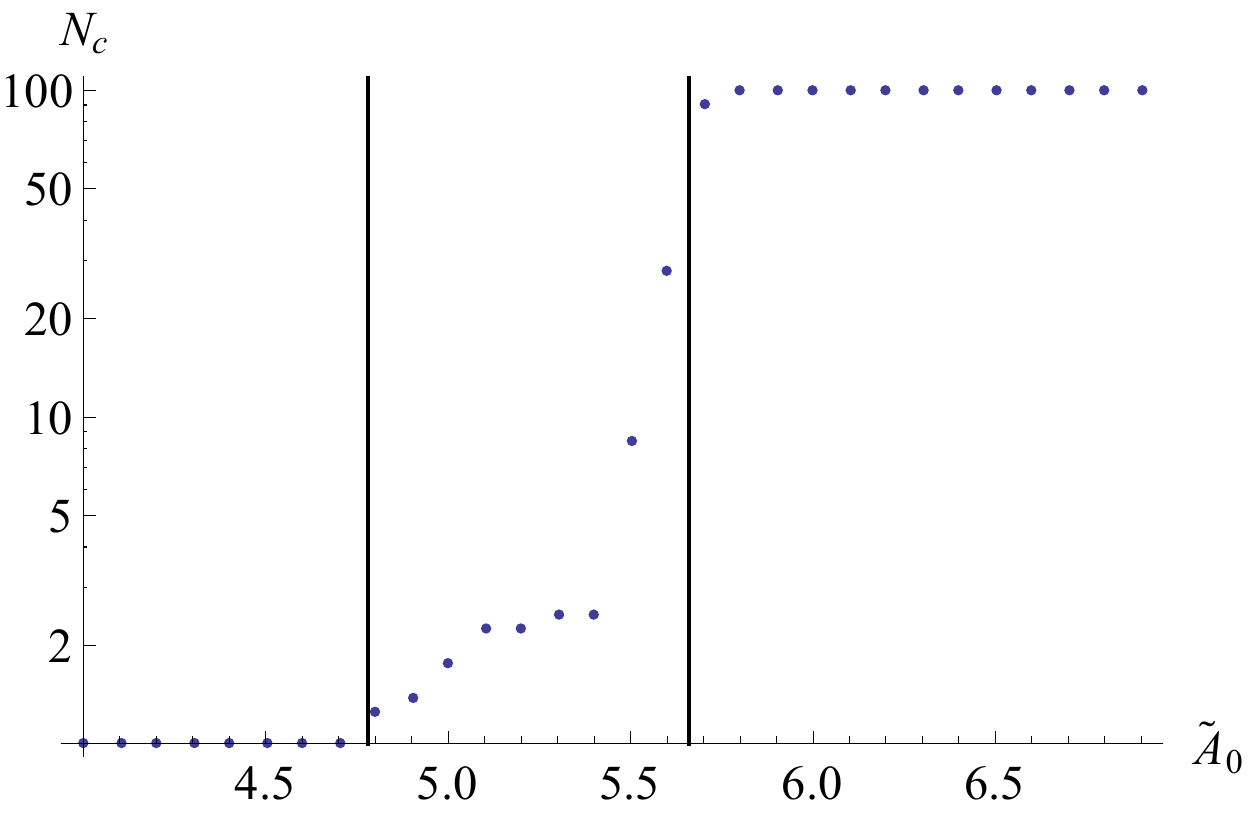}
\caption{The number of clusters, $N_{c}$, in a system with $N=100$ cells for $\tilde{K}=1$ and $\tilde{P}_{0}=3$. The two black lines designate the theoretical transition points between the Confluent and Cluster phases ($\tilde{A}_{0}=4.78$) and between the Cluster and Gas phases ($\tilde{A}_{0}=5.66$). Symbols are averages over $8$ simulations.}
\label{fig_conf_gas}
\end{figure}

\begin{figure}
\includegraphics[width=0.9\columnwidth]{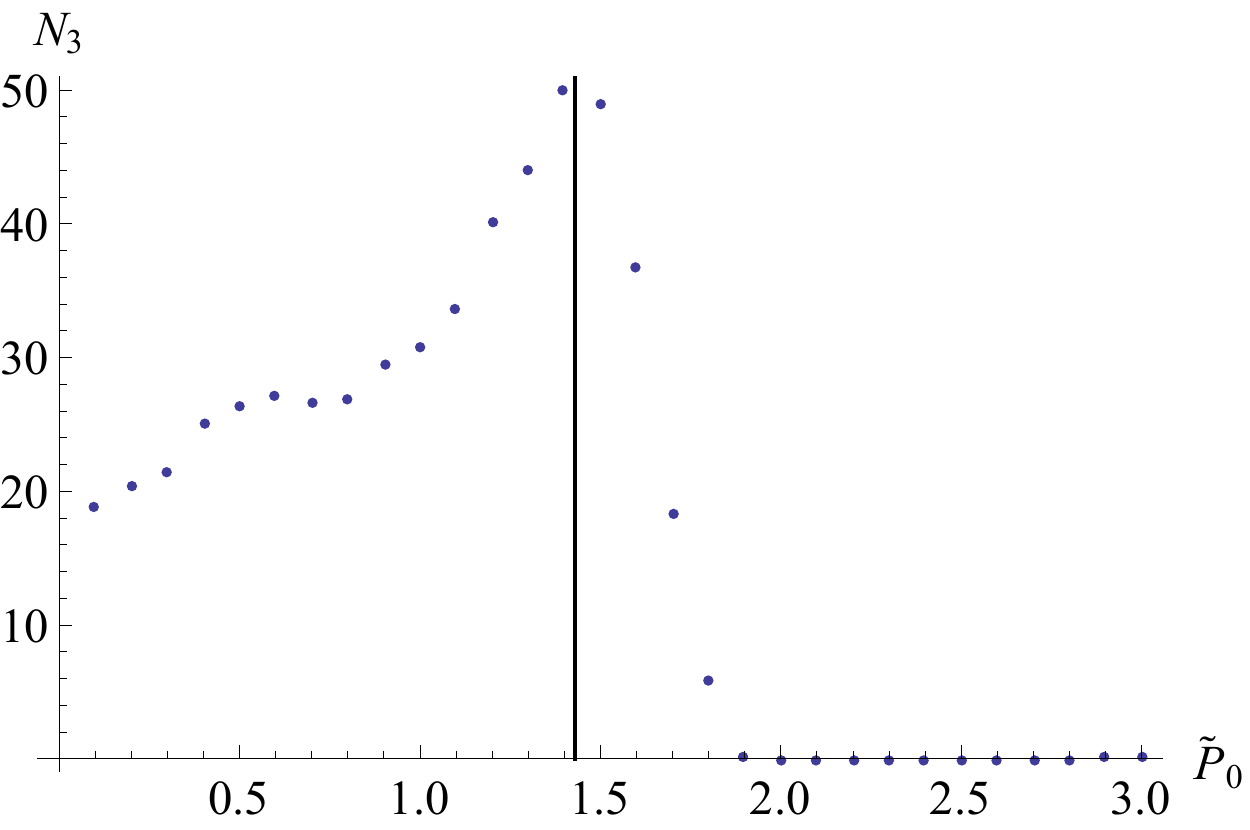}
\caption{The number of vertices connecting three cells, $N_{3}$, as a function of $\tilde{P}_{0}$ for $\tilde{K}=1$ and $\tilde{A}_{0}=8$. The black line shows the theoretical transition point between the Non-Confluent phase ($\tilde{P}_{0}\lessapprox 1.43$) and the Hexagonal phase ($\tilde{P}_{0}\gtrapprox 1.43$). Symbols are averages over $60$ simulations.}
\label{fig_nc_hex}
\end{figure}

The Confluent-Cluster, Cluster-Gas and Hexagonal-Gas transitions can be quantified by looking at the number of clusters which grows from $1$ in the Confluent and Hexagonal phases, to $N$ in the Gas phase, and is large and parameter dependent in the Cluster phase. Figure \ref{fig_conf_gas} shows how the number of clusters grows in the transition from the Confluent phase to the Gas phase. The transition between the Non-Confluent and the Hexagonal phases can be quantified via examining the number of vertices connecting $3$ or more cells (finite in the Non-Confluent phase and zero in the Hexagonal phases), as shown in Fig. \ref{fig_nc_hex}.

\subsection{The phase diagram in the dynamical $\ell$ model}

\begin{figure}
\includegraphics[width=\columnwidth]{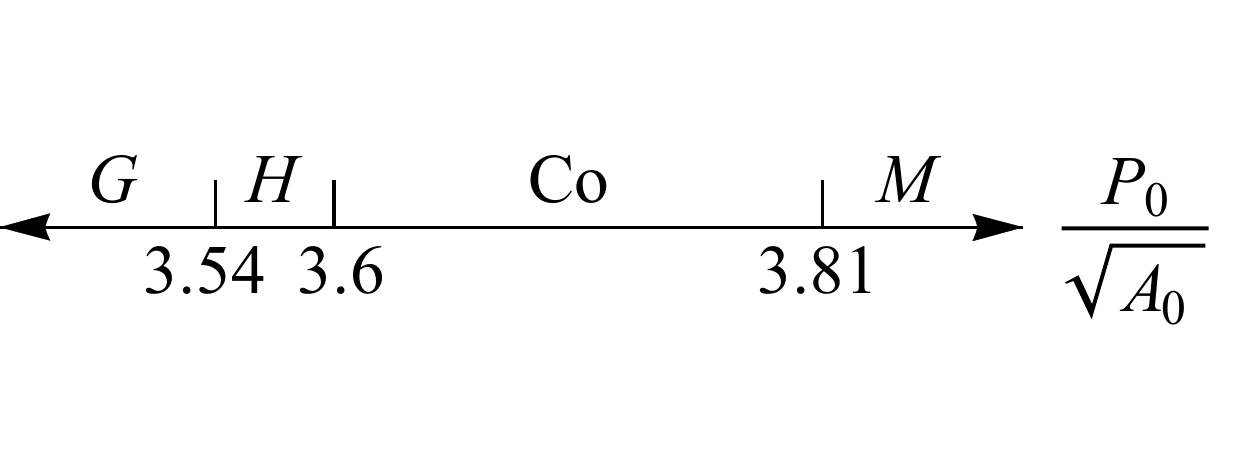}
\caption{The phase diagram of the dynamical $\ell$ model, which depends only on the value of the shape parameter $P_{0}/\sqrt{A_{0}}$. The letters corresponds to the Gas, Hexagonal, Confluent, and Minimal phases.}
\label{dyn_phase}
\end{figure}

In the dynamical $\ell$ model, in which the maximum radius of the cells can change in order to minimize the energy, the phase diagram is much simpler. For the two-cell and three-cell interactions, we combine Eqs. (\ref{zerol}), (\ref{2cellint}) and (\ref{3cellint}), and find that the two-cell interaction is attractive if $P_{0}/\sqrt{A_{0}}>2\sqrt{\pi}\approx3.54$, and that the three-cell interaction is attractive if
\begin{align}
\frac{P_{0}}{\sqrt{A_{0}}}>\frac{2\left(3+2\pi\right)}{\sqrt{\frac{9\sqrt{3}}{2}+6\pi}}\approx3.6 .
\end{align}
In the region $P_{0}/\sqrt{A_{0}}<3.54$, both interactions are repulsive, and thus the system is in the Gas phase. For $3.54<P_{0}/\sqrt{A_{0}}<3.6$ the two cell interaction is attractive, while the three cell interaction is repulsive, hence the system is in the Hexagonal phase. In the region $3.6<P_{0}/\sqrt{A_{0}}<q_{5}\approx3.8$ the system is in the Confluent phase, due to the same geometric constraints that occur in all related models \cite{Staple2010}. Finally, in the region $P_{0}/\sqrt{A_{0}}>3.8$ the system is in the Minimal phase. The phase diagram of the dynamical $\ell$ model is illustrated in Fig. \ref{dyn_phase}.

\subsection{Generalization to other models}

We emphasize here that the boundaries between the different phases remain qualitatively the same irrespective of the specific energy functional chosen, since the phases and the boundaries between them can be described in general terms. To summarize our overall picture: The Gas phase occurs when both two-cell and three-cell interactions are repulsive. The Cluster phase occurs when the two-cell interaction is repulsive while the three-cell interaction is attractive. The Hexagonal phase occurs when the two-cell interaction is attractive while the three-cell interaction is repulsive. The Confluent and Non-Confluent phases occur when the local minimum is for each cell to have identical neighboring cells connected in a confluent or non-confluent tissue respectively. The Minimal phase arises whenever there is enough flexibility for the system to reach a perfect state with zero energy.

\section{Finite effective temperature}
\label{sec_thermal}

\begin{figure}
\centering
\subfigure[Cluster]{\includegraphics[width=0.4\columnwidth]{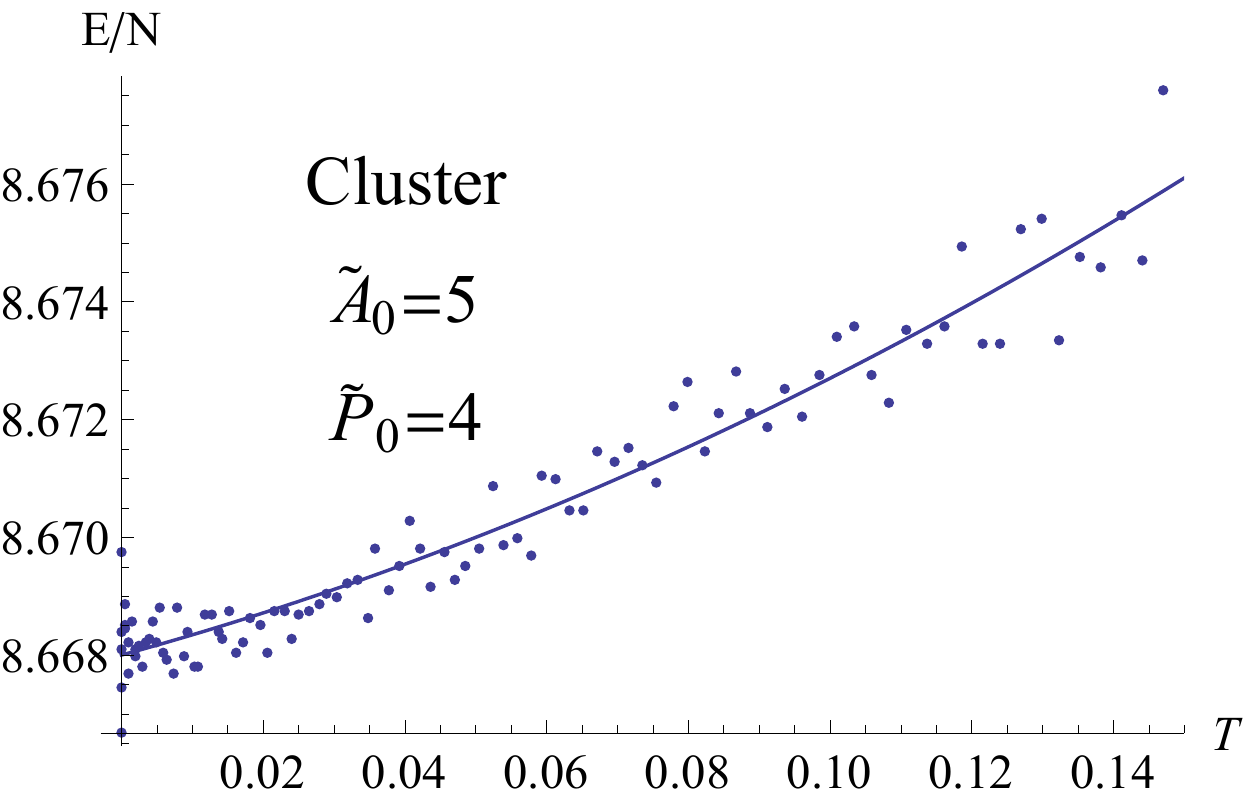}}
\subfigure[Hexagonal]{\includegraphics[width=0.4\columnwidth]{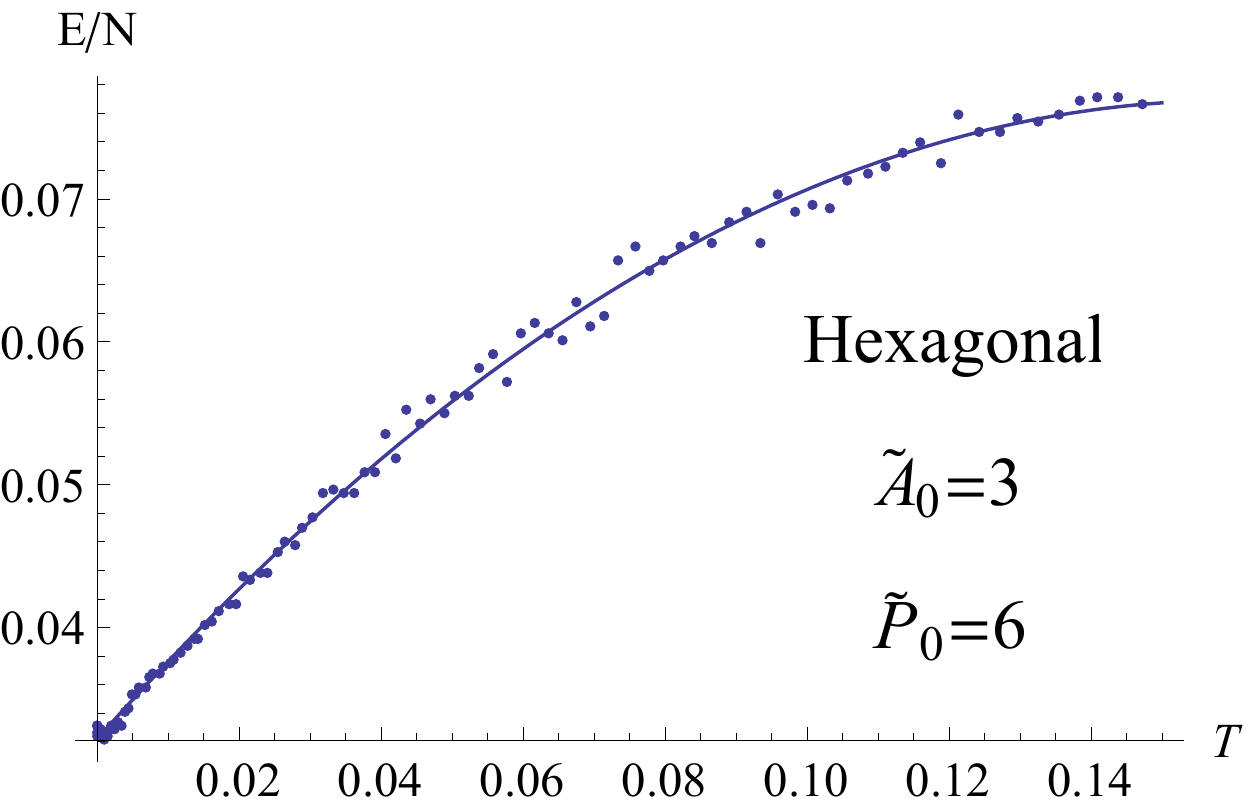}}\\
\subfigure[Minimal]{\includegraphics[width=0.4\columnwidth]{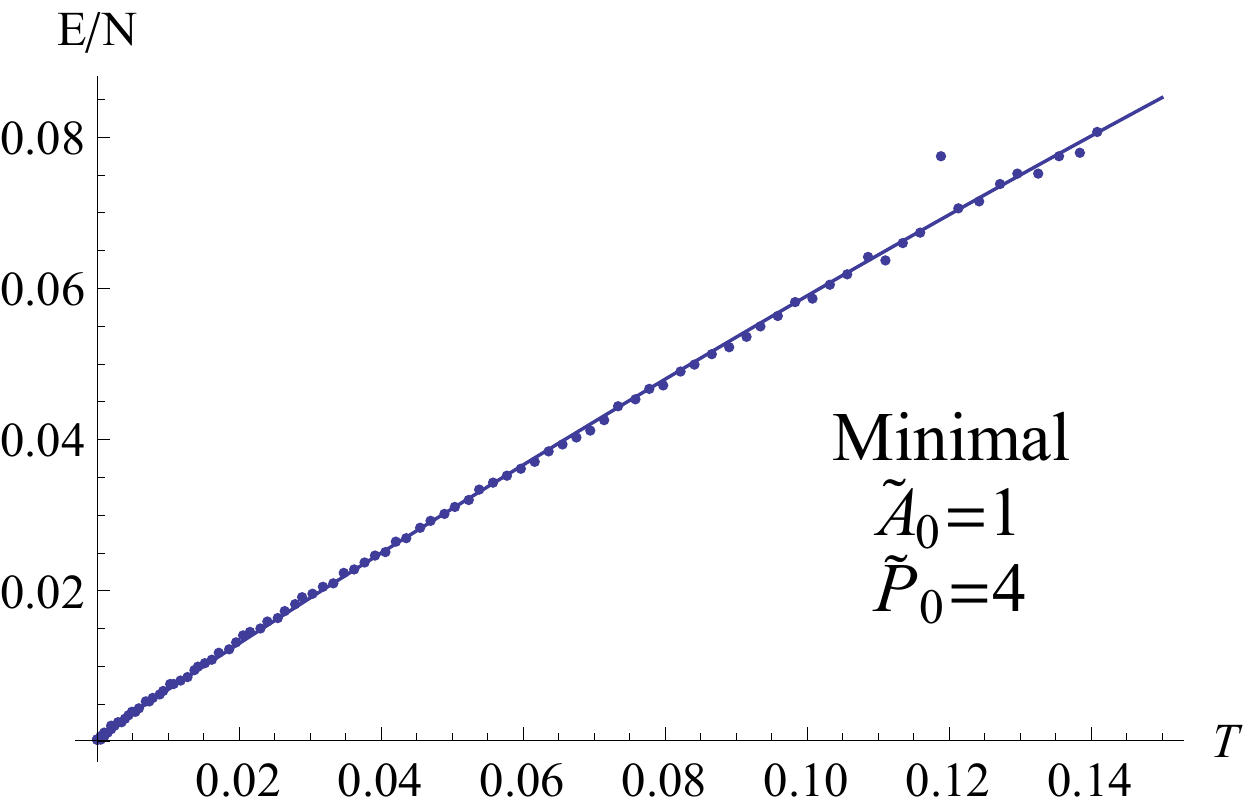}}
\subfigure[Non-Confluent]{\includegraphics[width=0.4\columnwidth]{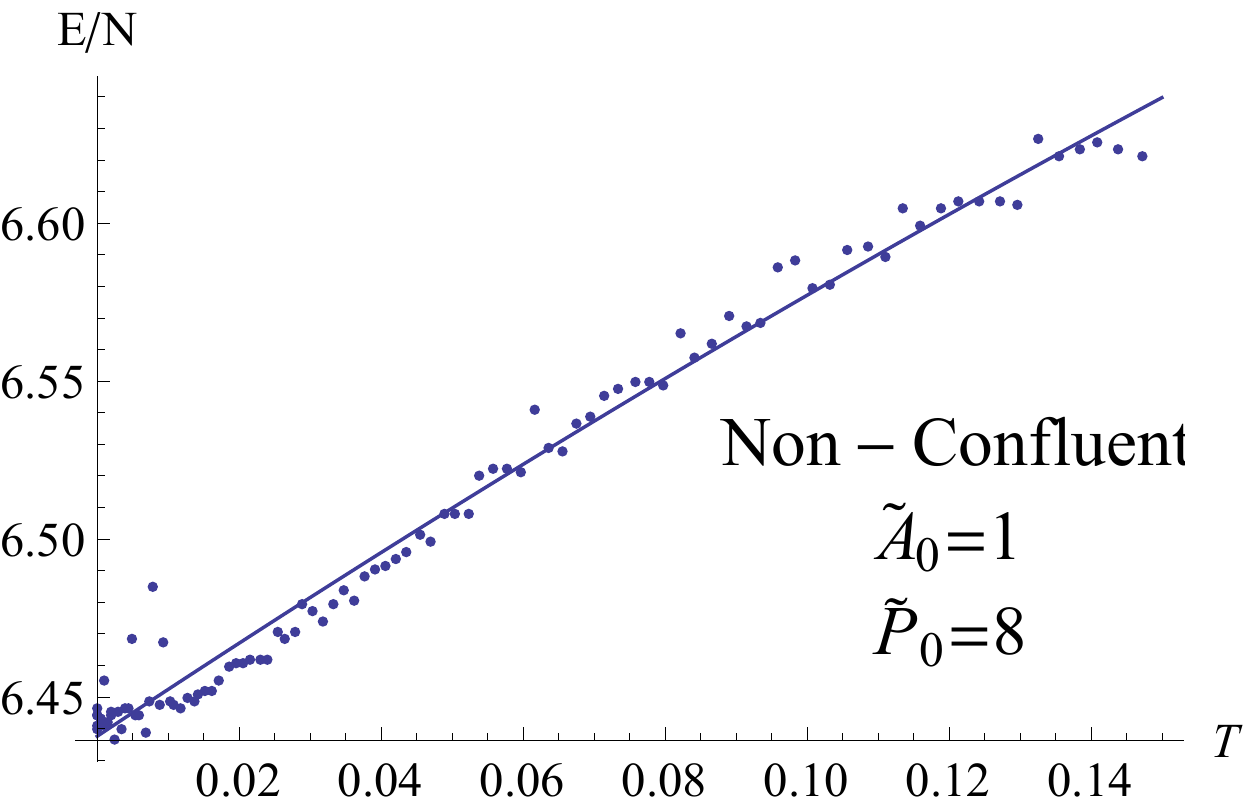}}\\
\caption{The mean energy per cell $E/N$ as a function of the temperature $T$ for a representative choice of parameters $\tilde{A}_{0}$ and $\tilde{P}_{0}$ for the Cluster, Hexagonal, Minimal and Non-Confluent phases. In all cases $\tilde{K}=1$. Each point in the plot represents the results of a single realization. The continuous lines are fits to a quadratic form.}
\label{fig_evt}
\end{figure}

The effect of the temperature on the phase diagram is rather weak. However, the temperature does give information about the different phases. By expanding the energy to second order in the deviation of the reference points from the global energy minimum obtained at $T=0$, we expect that at low temperatures the energy per cell, $\epsilon$, may be approximated by
\begin{align}
\epsilon(T)=\epsilon(T=0)+N_{f}T ,\label{et2}
\end{align}
where $N_{f}$ is the number of non-zero eigenvalues per particle of the Hessian matrix. See Section \ref{app_thermal} in the Appendix for the next order expansion of the energy. In all phases except the Confluent phase, we find that at $T=0$ the system indeed relaxes to the global energy minimum and Eq. (\ref{et2}) is valid. Simulation results for the mean energy per cell as a function of the temperature in the different phases is shown in Fig. \ref{fig_evt}. In all phases except the Confluent phase the energy increases with the temperature and agrees at low temperature with Eq. (\ref{et2}).

In the Gas and Cluster phases, we find that $N_{f}\approx0.1$. We discuss later why the Cluster phase behaves similarly to the Gas phase. Note that $N_{f}>0$ because the cells still interact occasionally. In the Minimal phase, $N_{f}\approx0.5\sim 0.8$, which shows the large freedom of movement the cells have within the aggregate. In the Hexagonal,  Confluent and Non-Confluent phases, we find that $N_{f}\approx1.3\sim 1.8$, which shows that in order for a cell to move it must overcome energy barriers. From these results we conclude that the Gas and Cluster phases behaves as a gas, the Minimal phase behaves as a liquid, while the Hexagonal and Non-Confluent phases behave as a solid. As will be shown below, the Confluent phase behaves as a glass.

\begin{figure}
\includegraphics[width=0.8\columnwidth]{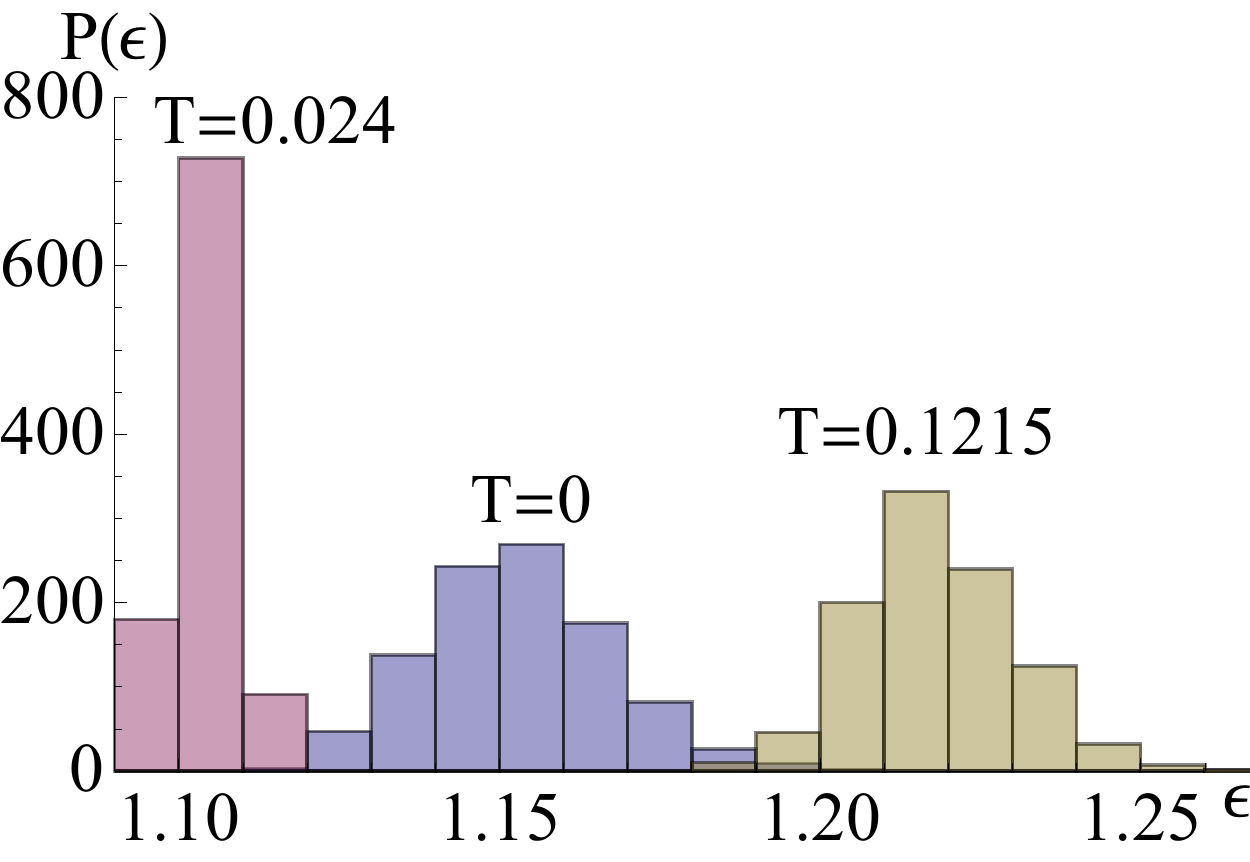}
\caption{Histogram of the energy minima the system relaxes to after a very long time in the Confluent phase for three different temperatures and a specific set of parameters ($\tilde{K}=1,\tilde{A}_{0}=2,\tilde{P}_{0}=3.5$). The data include $1000$ realizations. At $T=0$ the wide distribution shows the many energy minima the system can relax to. At $T\approx 0.024$ the mean energy is minimal for this set of parameters, and the distribution is very narrow. At $T=0.1215$ the wide distribution is due to the thermal fluctuations around the global minimum.}
\label{fig_e_histogram}
\end{figure}

\begin{figure}
\includegraphics[width=0.8\columnwidth]{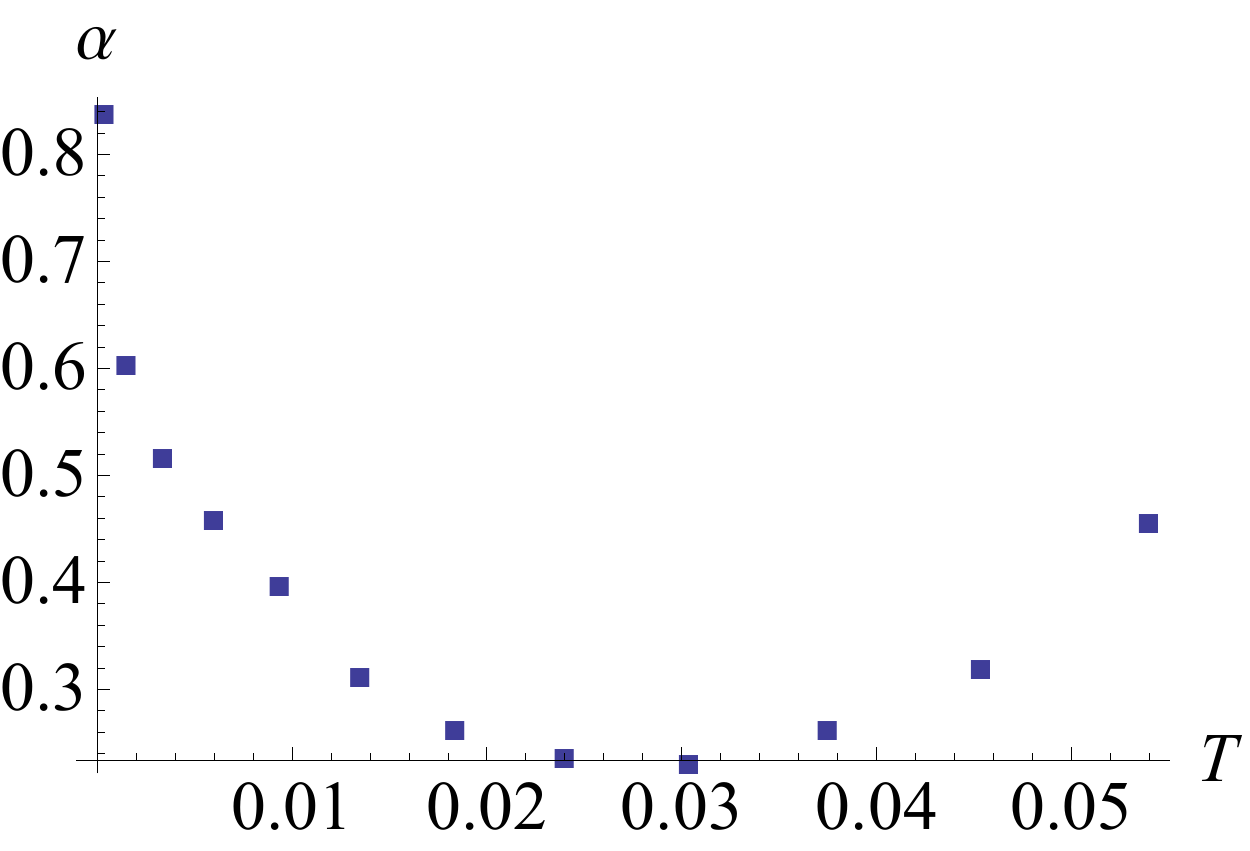}
\caption{The value of the exponent $\alpha$ describing the relaxation of the energy per cell with time, $\epsilon-\epsilon_{min}\propto t^{\alpha}$, as a function of the temperature for $\tilde{K}=1,\tilde{A}_{0}=2$ and $\tilde{P}_{0}=3.5$.}
\label{fig_alpha}
\end{figure}

In the Confluent phase, the system relaxes to only a local minimum of the energy, whose value depends on the initial conditions, as shown in Fig. \ref{fig_e_histogram}. 
As the temperature is increased, the mean energy actually decreases because the thermal fluctuations allow the system to overcome some of the energy barriers between those local minima and relax to different, lower minima. At a high enough temperature, the system behaves effectively as a thermal system fluctuating around the global minimum. Strictly speaking, this is not a phase transition because there is no discontinuity. Note that although at low temperatures each realization relaxes to a certain configuration, the mean energy per cell continues to relax very slowly to the global minimum as $\epsilon-\epsilon_{min}\propto t^{\alpha}$, with the dependence of $\alpha$ on the temperature shown in Fig. \ref{fig_alpha}. The value of the exponent $\alpha$ reaches a minimum of about $0.23$ at around $T\approx0.25$, which is the same temperature at which the mean energy is minimal.

In the Cluster phase, the temperature has a critical effect. At a high enough temperature, the three cell-contacts are broken by the thermal fluctuations. Because the two-cell interaction is repulsive, the cells cannot reconnect, and this causes the phase transition to the Gas phase. Although the thermal fluctuations also break the contacts in the other phases, the transition there should be continuous because the two-cell interaction is attractive in all the other phases (except for the Gas phase) and thus the cells can reconnect. We present in the SI two videos of a system in the Cluster phase ($\tilde{K}=1,\tilde{A}_{0}=4.5$ and $\tilde{P}_{0}=4.3$) with the same initial condition at different temperatures ($T=6\times 10^{-7}$ and $T=9.375\times 10^{-7}$). At the lower temperature, the cells mostly vibrate in place, while at the higher temperature, the clusters are broken and do not reconnect. Note that even the higher temperature is extremely small compared to the temperature range we consider elsewhere.

\begin{figure}
\subfigure[Cluster]{\includegraphics[width=0.4\columnwidth]{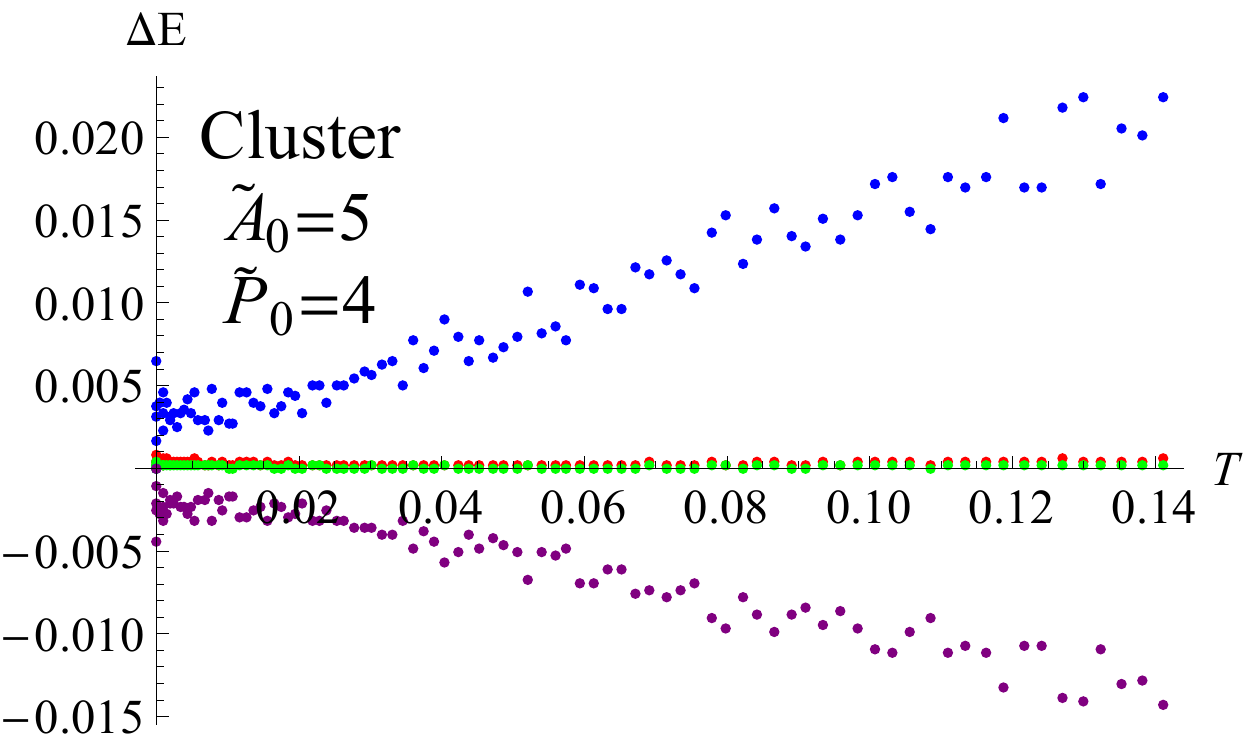}}
\subfigure[Hexagonal]{\includegraphics[width=0.4\columnwidth]{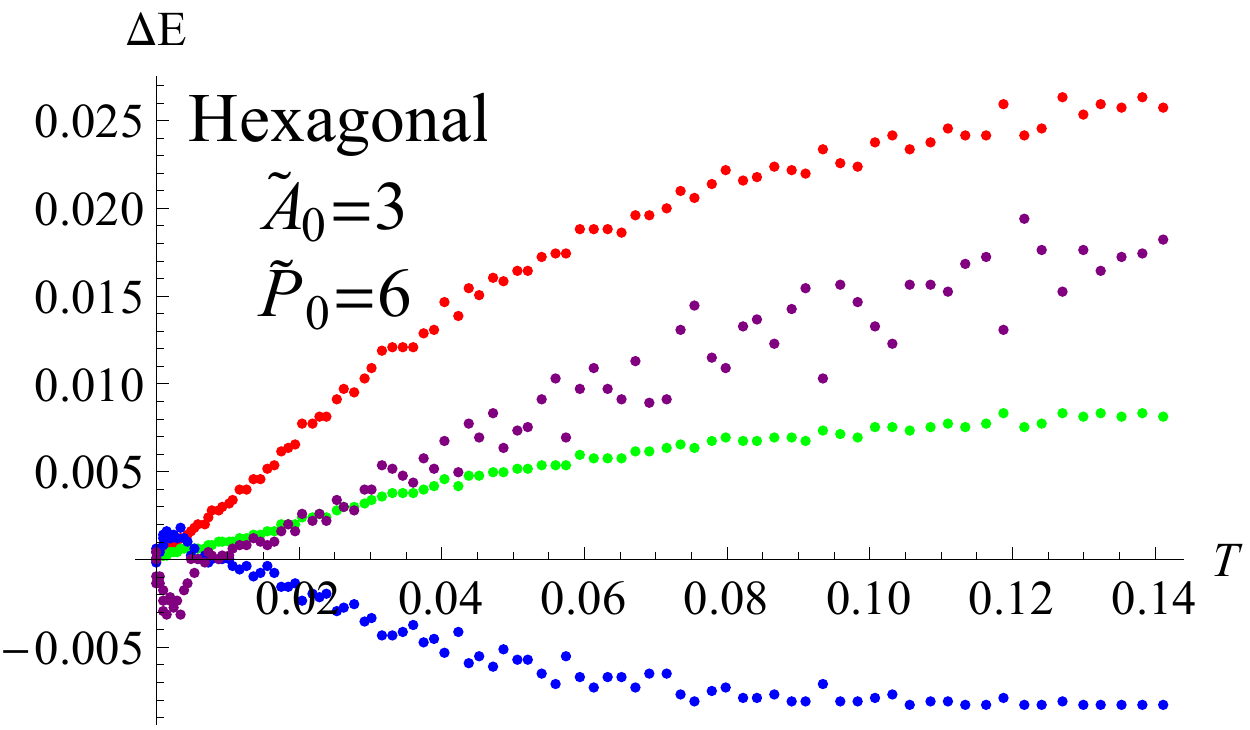}}\\
\subfigure[Minimal]{\includegraphics[width=0.4\columnwidth]{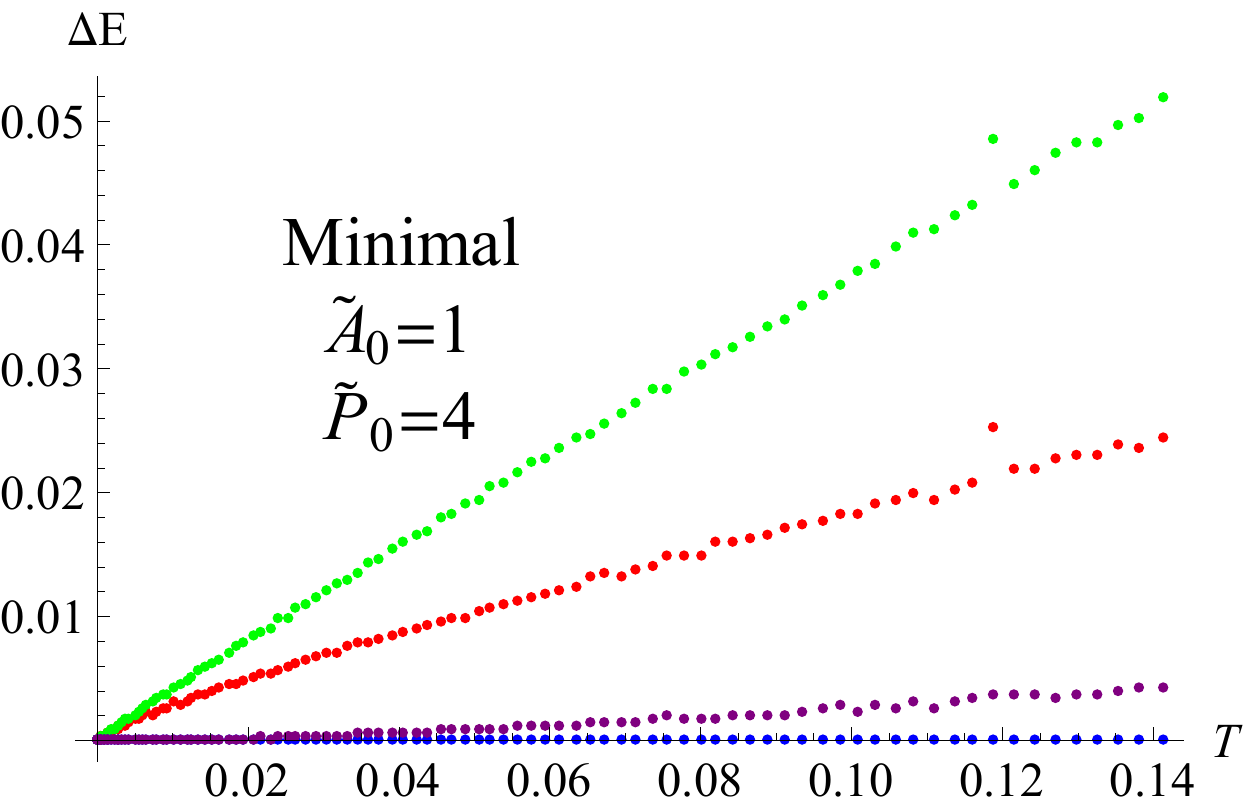}}
\subfigure[Non-Confluent]{\includegraphics[width=0.4\columnwidth]{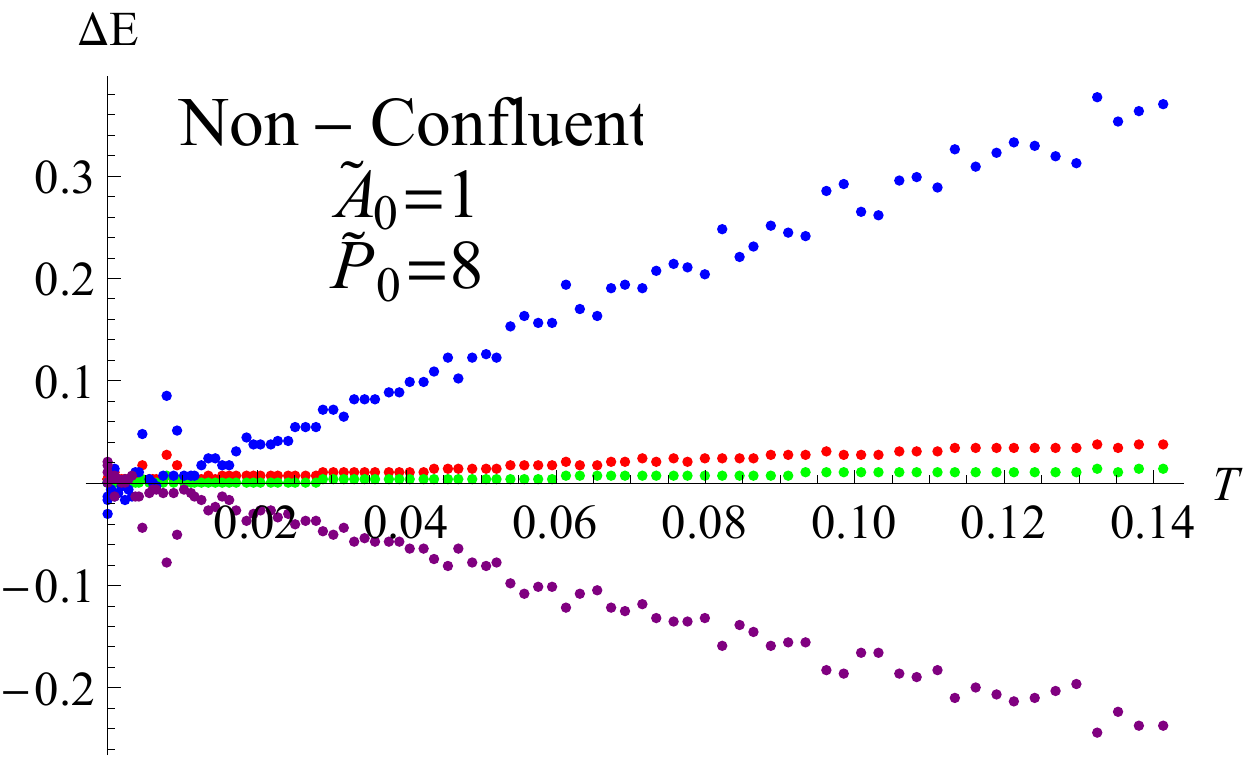}}\\
\caption{The deviation of each of the four parts of the energy, Eq. (\ref{4parts}), from its value at $T=0$ as a function of the temperature for a representative choice of parameters $\tilde{A}_{0}$ and $\tilde{P}_{0}$ for the Cluster, Hexagonal, Minimal and Non-Confluent phases. In all cases $\tilde{K}=1$. Each point in the plot represents the results of a single realization. Each color represents a different part: $\tilde{A}_{var}$ in red, $\tilde{P}_{var}$ in green, $\tilde{A}_{mean}$ in blue, and $\tilde{P}_{mean}$ in purple.}
\label{fig_de}
\end{figure}

In each of the phases there are different properties which are robust to the thermal fluctuations. In order to see them we divide the energy per cell, Eq. (\ref{energy1}) into four parts
\begin{align}
\tilde{\epsilon}=\tilde{A}_{var}+\tilde{A}_{mean}+\tilde{P}_{var}+\tilde{P}_{mean} ,
\end{align}
with
\begin{align}
&\tilde{A}_{var}=\langle\tilde{A}^{2}\rangle-\langle\tilde{A}\rangle^{2} ,
&\tilde{P}_{var}=\tilde{K}\left(\langle\tilde{P}^{2}\rangle-\langle\tilde{P}\rangle^{2}\right) ,\nonumber\\
&\tilde{A}_{mean}=\left(\langle\tilde{A}\rangle-\tilde{A}_{0}\right)^{2} ,
&\tilde{P}_{mean}=\tilde{K}\left(\langle\tilde{P}\rangle-\tilde{P}_{0}\right)^{2} .
\label{4parts}
\end{align}
Figure \ref{fig_de} shows how each part depends on the temperature in the different phases. We find that far from the Minimal phase, the variance terms, $\tilde{A}_{var}$ and $\tilde{P}_{var}$, are not affected by the temperature. Only in the Minimal phase can these two terms be the dominant terms in the energy. We also find that only in the Confluent phase $\tilde{A}_{mean}$ decreases with temperature and $\tilde{P}_{mean}$ increases with temperature, while the opposite behavior is observed at the Cluster, Non-Confluent, and Hexagonal phases, namely that $\tilde{A}_{mean}$ increases with temperature and $\tilde{P}_{mean}$ decreases. Geometrically, this means that in the Confluent phase increasing the temperature causes the cells to elongate, while in the Cluster, Non-Confluent, and Hexagonal phases increasing the temperature evidently causes the cells to become more compact.

\section{Summary}

In this paper, we investigated a Voronoi-based cellular model at thermal equilibrium, in which the total energy of the system is given by the area and circumference of each cell and their distance from a global preferred area and circumference. By requiring that each cell lies completely within a distance $\ell$ of its reference point, the model allows for non-confluent as well as confluent configurations. 
In open boundary conditions, we found six different phases, which occur due to competition between several factors: two-cell interactions, three-cell interactions, and geometric constraints. Previous studies, which constrained the cells to form a confluent tissue, identified two of these phases. All the other transitions cannot occur when the tissue is forced to be confluent. Furthermore, we argue that these phases are generic to open boundary models, regardless of the specific choice of energy functional. We expect that these transitions will help us understand the possible behaviors of cells in wound healing and in cancer metastasis, both cases where the tissue necessarily has an outer interface.

We approximated the active motion of cells as fluctuations induced by an effective temperature. Although this is a very crude approximation which cannot account for organized cell motion, amongst other things, our main focus in the paper is on the phase diagram of the system without activity. This approximation, however, allows us to probe two aspects of the system. First, we analyzed the Hessian and found that the Minimal phase is liquid-like, since the Hessian has many zero eigenvalues, while the Hexagonal and Non-Confluent phases are solid-like. Additionally, we investigated the energy landscape of the Confluent phase, and showed that it corresponds to glassy systems with many metastable states.

In a way, our work bridges the gap between several related works. The Confluent and Minimal phases are the two phases identified in works in which the confluency of the tissue is imposed, especially in \cite{Barton2017} which considers a confluent tissue under open boundary conditions. A more recent work considers deformable repulsive cells under compression \cite{Boromand2018}, which complements the Gas and Confluent phases and exhibits another way (compression) to achieve a phase transition.

The transition between the Confluent and Minimal phases has been seen in many experimental systems. The Gas phase appears when the cells are fully repulsive and detach from each other, see for example \cite{Zeng2016}. To our knowledge, the other three phases (Cluster, Non-Confluent, and  Hexagonal) have not been identified in experiment, perhaps in part since they had not been characterized as being distinct from the Confluent, Minimal and Gas phases. We note that in the Non-Confluent phase there are many almost degenerate pairs of three-cell vertices which effectively appear to be four-cell vertices, similar to what is seen in \cite{Kaliman2016}. The Cluster phase, as we have shown, is extremely sensitive to fluctuations, and thus dynamical measurements are needed to identify it. It general it may be simpler to identify the new phases from their dynamical behavior rather than their structure, especially in the presence of noise.
We also note that the difference between interior cells, with their polygonal shape, and the outer cells, with their rounded outer-facing boundaries, is readily apparent in typical pictures of tissues (see e.g. Fig. 4h from \cite{marieb}).

In this study we considered only two-dimensional systems. It would be interesting to expand our work to three-dimensional systems. We expect the phase diagram to be richer, since in three-dimensions the basic vertex in a confluent aggregate connects four cells, which adds another line to the phase diagram and may divide the existing phases further.
Other ways to extend this study includes adding self propulsion to the cells, considering cells of different phenotypes with different interactions, and considering elliptically-bounded cells.

\acknowledgements{ET and DAK acknowledge the support of the United States-Israel Binational Science Foundation, Grant no. 2015619. HL acknowledges the support of the NSF grant no. PHY-1605817. }

\bibliography{references}

\appendix
\section{Simulation details}
In the numerical simulations, we used the Euler method to advance the evolution equation, Eq. (\ref{eveq}), for $T=0$, and the Euler-Maruyama method for $T>0$. We set the time step to $\Delta t=0.001$. After each time step we first produced the Voronoi diagram using the plane-sweep algorithm \cite{Sweepline}, and then adjusted it to account for the finite radius of the cells. At $t=0$ the cells are randomly placed in a square of size $\ell\times\ell$.

The time until the system equilibrates varies, depending on the phase. In the Minimal, Cluster, Hexagonal, and Non-Confluent phases, the equilibration is fast at about $1-10$ time units (i.e. $10^{3}-10^{4}$ time steps). In the Gas phase, the system nearly equilibrates after $10^{3}$ time steps, but there are small two-cell repulsive contacts that can prolong the last steps of the equilibration by two or more orders of magnitude. In most cases, we did not wait for final equilibration in the Gas phase. In the Confluent phase, the equilibration is very slow, since the system is in a glassy state. Depending on how deep inside the Confluent phase the system is, the equilibration can take from $10^{4}$ to $10^{8}$ time steps, or even more.

\section{Calculation of the forces}
\label{app_forces}
In the simulations, we randomly generate the initial locations of the reference points of the $N$ cells and let them relax to a stable configuration. In order to do this we calculate the forces acting on each cell according to
\begin{align}
\vec{f}_{i}=-\frac{\partial\tilde{E}}{\partial\vec{r}_{i}} ,
\end{align}
where $\vec{r}_{i}$ is the location of the reference point of the cell. We assume overdamped dynamics such that the velocity of each cell is proportional to the force acting on it.

Under open boundary conditions, the shape of the cells is not determined solely by a Voronoi tessellation of the plane, as this would generate cells with infinite area. The cells have a maximum radius $\ell$, such that each cell contains all the points which are closest to its reference point, $\vec{r}_{i}$, and are at most a distance of $\ell$ from it \cite{Bock2010}.

The location of a vertex connecting three cells $i,j$ and $k$, $\vec{h}^{in}$, is given by
\begin{align}
&\vec{h}^{in}=\alpha^{jk}_{i}\vec{r}_{i}+\alpha^{ki}_{j}\vec{r}_{j}+\alpha^{ij}_{k}\vec{r}_{k} ,\label{eq_hin}
\end{align}
where
\begin{align}
\alpha^{jk}_{i}=\frac{\left|\vec{r}_{j}-\vec{r}_{k}\right|^2\left(\vec{r}_{i}-\vec{r}_{j}\right)\cdot\left(\vec{r}_{i}-\vec{r}_{k}\right)}{2\left|\left(\vec{r}_{i}-\vec{r}_{j}\right)\times\left(\vec{r}_{j}-\vec{r}_{k}\right)\right|^{2}} .\label{eq_alpha_def}
\end{align}
An outer vertex connecting two cells $i$ and $j$ is located at one of the two points which are at a distance $\ell$ from both $\vec{r}_{i}$ and $\vec{r}_{j}$. It is thus located at one of the two points
\begin{align}
&\vec{h}^{out}=\frac{\vec{r}_{i}+\vec{r}_{j}}{2}\pm\frac{\sqrt{4\ell^{2}-\left|\vec{r}_{i}-\vec{r}_{j}\right|^{2}}}{2\left|\vec{r}_{i}-\vec{r}_{j}\right|}\left(\vec{r}_{i}-\vec{r}_{j}\right)\times\hat{z} .\label{hout}
\end{align}
In order to calculate the perimeter and area of cell $i$, we enumerate the vertices surrounding it clockwise. The area and perimeter are given by
\begin{align}
&P_{i}=\sum_{m}P_{m,i} ,\nonumber\\
&A_{i}=\sum_{m}A_{m,i} .\label{eq_sum_ap}
\end{align}
where the summation variable indicates the vertices surrounding cell $i$. If at least one of the two vertices $m$ and $m+1$ is an inner vertex, then $P_{m,i}=\left|\vec{h}_{m}-\vec{h}_{m+1}\right|$, 
and
\begin{align}
A_{m,i}=\frac{1}{2}\left|\vec{h}_{m}\times\vec{h}_{m+1}+\vec{r}_{i}\times\left(\vec{h}_{m}-\vec{h}_{m+1}\right)\right| .\label{eq_ami}
\end{align}
If both vertices $m$ and $m+1$ are outer vertices, than $P_{m,i}=\ell\theta_{m}$ and $A_{m,i}=\frac{\ell^{2}\theta_{m}}{2}$, where
\begin{align}
\theta_{m}=\left\{\begin{array}{lr}
\theta^{(0)}_{m}&\theta^{(0)}_{m}<\pi\\
2\pi-\theta^{(0)}_{m}&\theta^{(0)}_{m}>\pi
\end{array}\right.
\end{align}
with
\begin{align}
\theta^{(0)}_{m}=\cos^{-1}\left[\frac{\left(\vec{h}_{m}-\vec{r}_{i}\right)\cdot\left(\vec{h}_{m+1}-\vec{r}_{i}\right)}{\ell^{2}}\right] .\label{eq_theta0}
\end{align}

The force acting on cell $i$ in direction $x$ may therefore be written as
\begin{align}
f_{i,x}=-\sum_{m}\frac{\partial E}{\partial\vec{h}_{m}}\cdot\frac{\partial\vec{h}_{m}}{\partial x_{i}}-\frac{\partial E}{\partial x_{i}} ,\label{eq_force}
\end{align}
where the sum is over all vertices connected to cell $i$, and $x_{i}$ is the $x$ coordinate of cell $i$. We obtain an analogous expression for the force acting in the $y$ direction.

The derivatives $\frac{\partial\vec{h}_{m}}{\partial x_{i}}$ depend on whether the vertex is an inner vertex or an outer vertex. For inner vertices we use Eq. (\ref{eq_hin}) and find that
\begin{align}
\frac{\partial\vec{h}^{in}}{\partial x_{i}}=\alpha^{j,k}_{i}\hat{x}+\frac{\partial\alpha^{i,k}_{j}}{\partial x_{i}}\vec{r}_{j}+\frac{\partial\alpha^{i,j}_{k}}{\partial x_{i}}\vec{r}_{k} ,
\end{align}
where $j$ and $k$ are the other two cells connected to the vertex. Using Eq. (\ref{eq_alpha_def}), we find that
\begin{align}
\frac{\partial\alpha^{i,k}_{j}}{\partial \vec{r}_{i}}=&\alpha^{i,k}_{j}\left[\frac{\vec{r}_{k}-\vec{r}_{j}}{\left(\vec{r}_{j}-\vec{r}_{i}\right)\cdot\left(\vec{r}_{j}-\vec{r}_{k}\right)}+2\frac{\vec{r}_{i}-\vec{r}_{k}}{\left|\vec{r}_{i}-\vec{r}_{k}\right|^{2}}\right.\nonumber\\
&\left.-2\frac{\left(\vec{r}_{k}-\vec{r}_{j}\right)\times\hat{z}}{\left|\left(\vec{r}_{i}-\vec{r}_{j}\right)\times\left(\vec{r}_{j}-\vec{r}_{k}\right)\right|}\right] .
\end{align}
For the outer vertices, we use Eq. (\ref{hout}) and find that
\begin{align}
\frac{\partial\vec{h}^{out}}{\partial x_{i}}=&\frac{\hat{x}}{2}\mp\frac{2\ell^{2}\left(x_{i}-x_{j}\right)\left[\left(y_{i}-y_{j}\right)\hat{x}-\left(x_{i}-x_{j}\right)\hat{y}\right]}{\sqrt{4\ell^{2}-\left|\vec{r}_{i}-\vec{r}_{j}\right|^{2}}\left|\vec{r}_{i}-\vec{r}_{j}\right|^{3}}\nonumber\\
&\mp\sqrt{\frac{\ell^{2}}{\left|\vec{r}_{i}-\vec{r}_{j}\right|^{2}}-\frac{1}{4}}\hat{y} .
\end{align}

The derivative of the energy with respect to the location of the vertices is
\begin{align}
\frac{\partial E}{\partial\vec{h}}=2\sum_{i}K_{A}\left(A_{i}-A_{0}\right)\frac{\partial A_{i}}{\partial\vec{h}}+K_{P}\left(P_{i}-P_{0}\right)\frac{\partial P_{i}}{\partial\vec{h}} ,
\end{align}
where the sum is over the cells connected to the vertex.
Using Eq. (\ref{eq_sum_ap}), the derivative of the area with respect to $\vec{h}_{m}$ is
\begin{align}
\frac{\partial A_{i}}{\partial\vec{h}_{m}}=\frac{\partial A_{m,i}}{\partial\vec{h}_{m}}+\frac{\partial A_{m-1,i}}{\partial\vec{h}_{m}} ,
\end{align}
and similarly for the circumference. For an edge connecting two cells these derivatives are
\begin{align}
\frac{\partial A_{m,i}}{\partial\vec{h}_{m}}=&\frac{\left(\vec{h}_{m+1}-\vec{r}_{i}\right)\times\hat{z}}{2}\nonumber\\
&sgn\left[\hat{z}\cdot\left(\vec{h}_{m}\times\vec{h}_{m+1}+\vec{r}_{i}\cdot\left(\vec{h}_{m}-\vec{h}_{m+1}\right)\right)\right] ,\nonumber\\
\frac{\partial A_{m-1,i}}{\partial\vec{h}_{m}}=&\frac{\left(\vec{h}_{m-1}-\vec{r}_{i}\right)\times\hat{z}}{2}\nonumber\\
&sgn\left[\hat{z}\cdot\left(\vec{h}_{m}\times\vec{h}_{m-1}+\vec{r}_{i}\cdot\left(\vec{h}_{m}-\vec{h}_{m-1}\right)\right)\right] ,\nonumber\\
\frac{\partial P_{m,i}}{\partial\vec{h}_{m}}=&\frac{\vec{h}_{m}-\vec{h}_{m+1}}{\left|\vec{h}_{m}-\vec{h}_{m+1}\right|} ,\nonumber\\
\frac{\partial P_{m-1,i}}{\partial\vec{h}_{m}}=&\frac{\vec{h}_{m}-\vec{h}_{m-1}}{\left|\vec{h}_{m}-\vec{h}_{m-1}\right|}
\end{align}
For an edge connecting a cell to the unoccupied region, we use Eq. (\ref{eq_theta0}) and find that
\begin{align}
\frac{\partial\theta^{(0)}_{m}}{\partial\vec{h}_{m}}=-\frac{\vec{h}_{m+1}-\vec{r}_{i}}{\sin\theta^{(0)}_{m}} .
\end{align}

The only place in which the explicit derivative of $E$ with respect to $\vec{r}_{i}$ in Eq. (\ref{eq_force}) does not vanish is at the boundaries between a cell and the unoccupied region, the contribution from which is straightforward to evaluate by using
\begin{align}
\frac{\partial\theta^{(0)}_{m}}{\partial\vec{r}_{i}}=\frac{\vec{h}_{m}+\vec{h}_{m+1}-2\vec{r}_{i}}{\sin\theta^{(0)}_{m}} .
\end{align}

\section{Stability of five cell configuration inside the Gas phase}
\label{5cell}
Within the Gas phase, where both the two-cell and three-cell interaction are repulsive, there are several stable configurations. Here we analyze one of them.

\begin{figure}
\centering
\includegraphics[width=0.3\columnwidth]{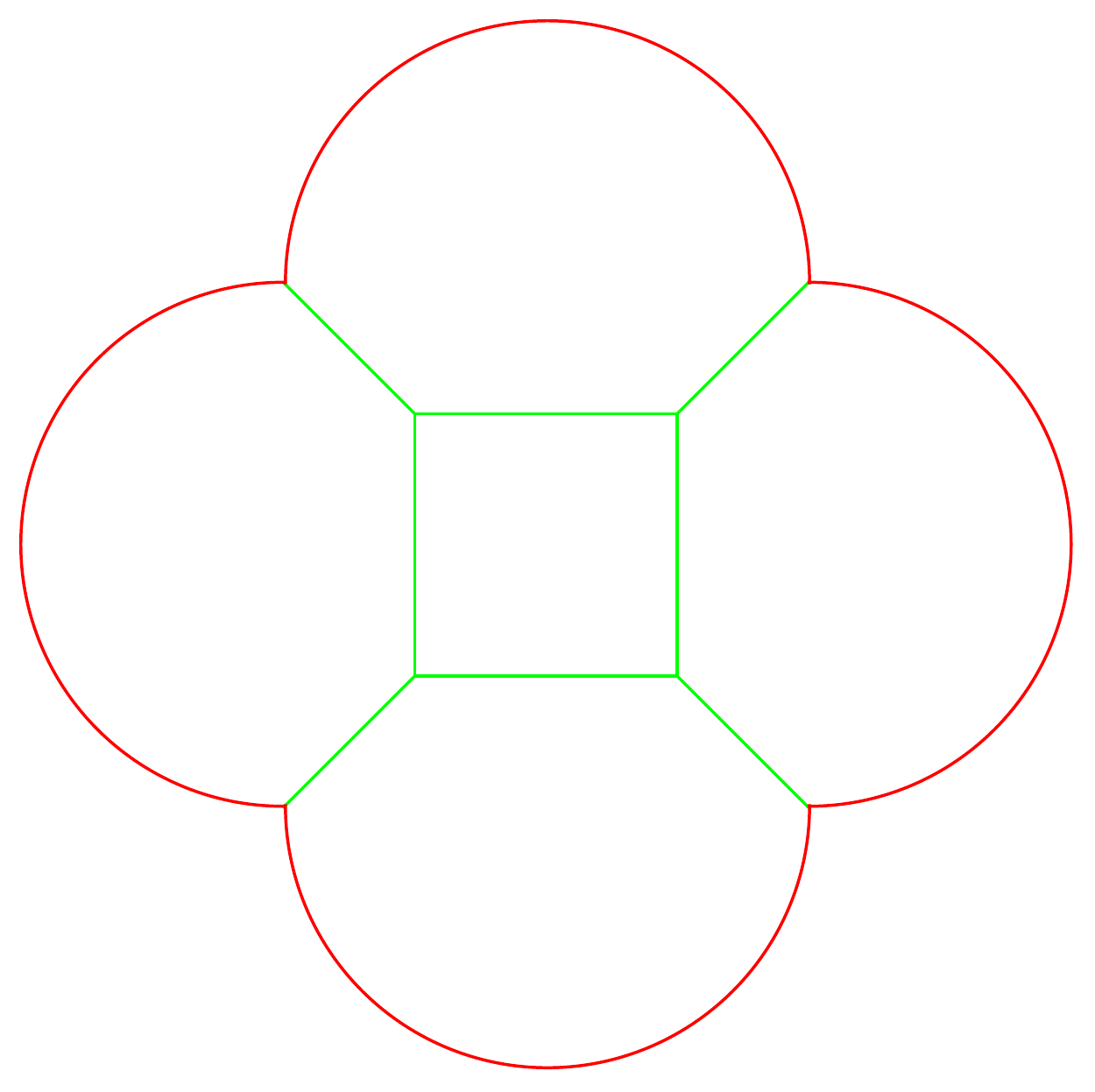}
\caption{An example of a rare five-cell structure which is stable even when three-cell interactions are repulsive.}
\label{fig_penta}
\end{figure}

Consider the five-cell structure shown in Fig. \ref{fig_penta}. 
The central cell is located at the origin, and the four outer cells are located at $\vec{r}_{1}=R\hat{x}, \vec{r}_{2}=R\hat{y}, \vec{r}_{3}=-R\hat{x}$ and $\vec{r}_{4}=-R\hat{y}$. The area and circumference of the central cell are
\begin{align}
&A_{cen}=R^{2} ,\nonumber\\
&P_{cen}=4R ,
\end{align}
and the area and circumference of the four outer cells are
\begin{align}
&A_{out}=\frac{1}{4}\left[R\left(R+2\sqrt{2\ell^{2}-R^{2}}\right)+2\ell^{2}\theta_{out}+\right.\nonumber\\
&\left.+4\ell^{2}\left(\pi-\theta_{out}\right)H\left(R-\ell\right)\right] ,\nonumber\\
&P_{out}=R+\sqrt{4\ell^{2}-2R^{2}}+\ell\theta_{out}+2\ell\left(\pi-\theta_{out}\right)H\left(R-\ell\right) ,
\end{align}
where $H(z)$ is the Heaviside step function, and
\begin{align}
\theta_{out}=\cos^{-1}\left(-\frac{R\sqrt{2\ell^{2}-R^{2}}}{\ell^{2}}\right) .
\end{align}

The structure is stable when $\partial E/\partial R>0$ at $R=\sqrt{2}\ell$. Solving the inequality yields
\begin{align}
\sqrt{2}\left(10+3\pi-8\tilde{A}_{0}\right)+4\tilde{K}\left(10\sqrt{2}+3\pi-4\tilde{P}_{0}\right)>0 .
\end{align}
This critical line lies inside the Gas phase.

\section{Full expressions for analysing the stability of a honeycomb lattice}
\label{app_honeycomb}

The full expression of the matrix $\Omega$ appearing in Eq. (\ref{omegadef}) is 
\begin{align}
&\Omega_{x,x}=\frac{\tilde{P}^{2}\cos^{2}\left(\tilde{q}_{x}\right)\cos^{2}\left(\tilde{q}_{y}\right)}{6}-\nonumber\\
&-4\tilde{K}\left[3-\cos\left(2\tilde{q}_{x}\right)\cos^{2}\left(\tilde{q}_{y}\right)-7\sin^{2}\left(\tilde{q}_{y}\right)-\right.\nonumber\\
&\left.-2\cos\left(\tilde{q}_{x}\right)\cos\left(\tilde{q}_{y}\right)\right]-\nonumber\\
&-\frac{8\tilde{K}\tilde{P}_{0}}{\tilde{P}}\left[1+4\sin^{2}\left(\tilde{q}_{y}\right)-\cos\left(\tilde{q}_{x}\right)\cos\left(\tilde{q}_{y}\right)\right] ,\nonumber\\
&\Omega_{y,y}=\frac{\tilde{P}^{2}\sin^{2}\left(\tilde{q}_{y}\right)\left[\cos\left(\tilde{q}_{x}\right)+2\cos\left(\tilde{q}_{y}\right)\right]^{2}}{18}+\nonumber\\
&+\frac{4\tilde{K}}{3}\left\{20-2\cos^{2}\left(\tilde{q}_{x}\right)\cos^{2}\left(\tilde{q}_{y}\right)-\cos\left(4\tilde{q}_{y}\right)-\right.\nonumber\\
&\left.-2\cos\left(\tilde{q}_{x}\right)\left[8\cos\left(\tilde{q}_{y}\right)+\cos\left(3\tilde{q}_{y}\right)\right]\right\}-\nonumber\\
&-\frac{24\tilde{K}\tilde{P}_{0}}{\tilde{P}}\left[1-\cos\left(\tilde{q}_{x}\right)\cos\left(\tilde{q}_{y}\right)\right] ,\nonumber\\
&\Omega_{x,y}=\Omega_{y,x}=\nonumber\\
&=\frac{\tilde{P}^{2}}{3^{1/4}24}\sin\left(\tilde{q}_{x}\right)\sin\left(2\tilde{q}_{y}\right)\left[\cos\left(\tilde{q}_{x}\right)+2\cos\left(\tilde{q}_{y}\right)\right]-\nonumber\\
&-\frac{8\tilde{K}}{\sqrt{3}}\sin\left(\tilde{q}_{x}\right)\sin\left(\tilde{q}_{y}\right)\left[2-\cos\left(2\tilde{q}_{y}\right)-\right.\nonumber\\
&\left.-\cos\left(\tilde{q}_{x}\right)\cos\left(\tilde{q}_{y}\right)\right]+\nonumber\\
&+\frac{4\sqrt{3}\tilde{K}\tilde{P}_{0}}{\tilde{P}}\sin\left(\tilde{q}_{x}\right)\sin\left(\tilde{q}_{y}\right) ,
\end{align}
where we used for brevity
\begin{align}
&\tilde{q}_{x}=\frac{q_{x}P}{4} ,\nonumber\\
&\tilde{q}_{y}=\frac{q_{y}P}{4\sqrt{3}} ,
\end{align}
with $P$ being the circumference of each cell at equilibrium.

The full expression of the functions $g_{i}(\vec{q})$ appearing in Eq. (\ref{fung}) are
\begin{align}
&g_{1}(\vec{q})=42+4\cos\left(2\tilde{q}_{y}\right)-14\cos\left(4\tilde{q}_{y}\right)+4\cos\left(6\tilde{q}_{y}\right)+\nonumber\\
&+2\cos\left(3\tilde{q}_{x}\right)\left[5\cos\left(\tilde{q}_{y}\right)+4\cos\left(3\tilde{q}_{y}\right)\right]+\nonumber\\
&+2\cos\left(2\tilde{q}_{x}\right)\left[-9-14\cos\left(2\tilde{q}_{y}\right)+5\cos\left(4\tilde{q}_{y}\right)\right]+\nonumber\\
&+2\cos\left(\tilde{q}_{x}\right)\left[4\cos\left(\tilde{q}_{y}\right)-18\cos\left(3\tilde{q}_{y}\right)+5\cos\left(5\tilde{q}_{y}\right)\right] ,\nonumber\\
&g_{2}(\vec{q})=\frac{1}{243}\left\{129-52\cos\left(2\tilde{q}_{y}\right)-7\cos\left(4\tilde{q}_{y}\right)+\right.\nonumber\\
&\left.+2\cos\left(2\tilde{q}_{y}\right)+\cos\left(3\tilde{q}_{x}\right)\left[5\cos\left(\tilde{q}_{y}\right)+4\cos\left(3\tilde{q}_{y}\right)\right]+\right.\nonumber\\
&\left.+4\cos\left(2\tilde{q}_{x}\right)\left[2-5\cos\left(2\tilde{q}_{y}\right)\right]-\right.\nonumber\\
&\left.-\cos\left(\tilde{q}_{x}\right)\left[104\cos\left(\tilde{q}_{y}\right)-18\cos\left(3\tilde{q}_{y}\right)-5\cos\left(5\tilde{q}_{y}\right)\right]\right\} ,\nonumber\\
&g_{3}(\vec{q})=\frac{2}{27}\left\{\cos\left(2\tilde{q}_{x}\right)+\right.\nonumber\\
&\left.+\left[3-4\cos\left(\tilde{q}_{x}\right)\cos\left(\tilde{q}_{y}\right)\right]\left[2-\cos\left(2\tilde{q}_{y}\right)\right]\right\} .
\end{align}

\section{Different cell-cell and cell-boundary interface energy}
\label{dif_inter}
In this section we show how letting the cell-cell and cell-boundary interface energies differ affects the phase diagram. Specifically, we start from Eq. (\ref{energy0}) and set $A_{0,i}=A_{0}$ for all cells, $\Gamma_{i}=K_{P}$ for all cells, $\Lambda_{i,j}=-K_{P}P_{0}$ for cell-cell interfaces, and $\Lambda_{i,j}=-K_{P}P_{0}+\frac{\Delta}{2}$ for cell-boundary interfaces, such that up to a constant the energy may be written as
\begin{align}
E=&K_{A}\sum_{i}\left(A_{i}-A_{0}\right)^{2}+K_{P}\sum_{i}\left(P_{i}-P_{0}\right)^{2}-\nonumber\\
&-\Delta \sum_{i}P^{out}_{i} ,
\end{align}
where $P_{i}$ is the total circumference of the cells and $P^{out}_{i}$ is the part which is not in contact with any other cell.

Because the only difference a non-zero value of $\Delta$ can have is on the cells at the boundaries, it does not affect the transition between the Confluent and Minimal phases, and between the Minimal and Non-Confluent phases. The only effect it has is on the other transitions, which are determined by whether the two- and three-cell interactions are attractive or repulsive.

Now the two cell interaction is attractive if
\begin{align}
2\left(\pi-\tilde{A}_{0}\right)+\tilde{K}\left(2\pi-\tilde{P}_{0}\right)-2\pi\tilde{\Delta}>0 ,
\end{align}
and the three cell interaction is attractive if
\begin{align}
&\left(\frac{3\sqrt{3}+4\pi}{6}-\tilde{A}_{0}\right)+\nonumber\\
&+\frac{2}{3}\left(3-\sqrt{3}\right)\tilde{K}\left(\frac{2\left(3+2\pi\right)}{3}-\tilde{P}_{0}\right)-\nonumber\\
&-\frac{8\pi}{9}\left(3-\sqrt{3}\right)\tilde{\Delta}>0 ,
\end{align}
where
\begin{align}
\tilde{\Delta}=\frac{\Delta}{K_{A}\ell^{4}} .
\end{align}
Hence, in the $\tilde{P}_{0}-\tilde{A}_{0}$ plane, the transitions are still two non-parallel straight lines for any fixed value of $\tilde{\Delta}$. The only qualitative difference would be if the two lines intersect at a non-physical range of parameters, such that one of the phases will not be accessible. The only physical restriction on the parameters is $\tilde{K}>0$ and $\tilde{A}_{0}>0$, while $\tilde{P}_{0}$ and $\tilde{\Delta}$ can have any real value. The value of $\tilde{A}_{0}$ at the intersection between the two lines is
\begin{align}
&\tilde{A}_{int}=\frac{1}{66}\left[-9\left(4+3\sqrt{3}\right)+4\pi\left(21+2\sqrt{3}\right)+\right.\nonumber\\
&\left.+8\left(\pi-3\right)\left(5+\sqrt{3}\right)\tilde{K}-8\pi\left(5+\sqrt{3}\right)\tilde{\Delta}\right] .
\end{align}
This is negative for $\tilde{\Delta}>\tilde{\Delta}_{c}\left(\tilde{K}\right)$ where
\begin{align}
&\tilde{\Delta}_{c}\left(\tilde{K}\right)=\frac{4\pi\left(9-\sqrt{3}\right)-9\left(1+\sqrt{3}\right)}{16\pi}+\left(1-\frac{3}{\pi}\right)\tilde{K}\approx\nonumber\\
&\approx 1.3+0.045\tilde{K} .
\end{align}
Therefore, the Hexagonal phase is inaccessible for large enough values of $\tilde{\Delta}$ as long as $\tilde{K}$ is small enough.

\section{Expansion of the energy to second order in the temperature}
\label{app_thermal}
In this section we investigate the effects of low temperatures on the energy.
Assuming the system has achieved thermal equilibrium, the probability to find it in a certain configuration $C$ is given by the Boltzmann distribution,
\begin{align}
{\cal P}\left(C\right)=\frac{e^{-E\left(C\right)/T}}{\int e^{-E\left(C\right)/T}dC} ,
\end{align}
where the configuration $C$ encodes the location of all the cells. Assuming that at $T=0$ the system reaches its lowest possible energy, then for small $T$ we may expand the energy around that minimum,
\begin{align}
E\approx E_{0}+\frac{1}{2}\sum^{2N}_{i,j=1}x_{i}x_{j}U_{i,j}\equiv E_{0}+\frac{1}{2}\vec{x}^{T}U\vec{x} ,
\end{align}
where $x_{i}$ are the deviations of the $2N$ degrees of freedom of the $N$ cells from their locations in the minimum energy configuration. The mean energy per cell is thus
\begin{align}
&\frac{\left\langle E\right\rangle}{N}=\frac{1}{N}\frac{\int\left(E_{0}+\frac{1}{2}\vec{x}^{T}U\vec{x}\right)e^{-\left(E_{0}+\frac{1}{2}\vec{x}^{T}U\vec{x}\right)/T}d\vec{x}}{\int e^{-\left(E_{0}+\frac{1}{2}\vec{x}^{T}U\vec{x}\right)/T}d\vec{x}}=\nonumber\\
&=\frac{1}{N}\left(E_{0}+\frac{1}{2}\frac{\int\frac{1}{2}\vec{x}^{T}U\vec{x}e^{-\frac{1}{2}\vec{x}^{T}U\vec{x}/T}d\vec{x}}{\int e^{-\frac{1}{2}\vec{x}^{T}U\vec{x}/T}d\vec{x}}\right) .
\end{align}
Using a well known trick we can simplify the nominator

\begin{align}
\frac{\left\langle E\right\rangle}{N}=\frac{1}{N}\left(E_{0}+\frac{-T\frac{\partial}{\partial\alpha}\int e^{-\frac{1}{2}\alpha\vec{x}^{T}U\vec{x}/T}d\vec{x}}{\int e^{-\frac{1}{2}\vec{x}^{T}U\vec{x}/T}d\vec{x}}\right)\Big|_{\alpha=1} .
\end{align}
Evaluating the integrals yields
\begin{align}
\frac{\left\langle E\right\rangle}{N}=\frac{1}{N}\left(E_{0}+\frac{-T\frac{\partial}{\partial\alpha}\sqrt{\frac{\left(2\pi\right)^{2n}}{|\alpha U'/T|}}}{\sqrt{\frac{\left(2\pi\right)^{2n}}{|U'/T|}}}\right)\Big|_{\alpha=1} ,
\end{align}
where $U'$ is the matrix $U$ reduced to its non-zero eigenvalues, $2n$ is the size of $U'$, and $|U'|$ is the determinant of $U'$. Simplifying yields
\begin{align}
\frac{\left\langle E\right\rangle}{N}=\frac{1}{N}\left(E_{0}-T\frac{\partial}{\partial\alpha}\alpha^{-n}\right)\Big|_{\alpha=1} ,
\end{align}
Evaluating the derivative and setting $\alpha=1$ yields
\begin{align}
\frac{\left\langle E\right\rangle}{N}=\frac{1}{N}\left(E_{0}+nT\right) .\label{evsd}
\end{align}

In order to find the next order correction, we expand the energy to fourth order in $\vec{x}$ and assume that it can be written as
\begin{align}
E=E_{0}+\frac{1}{2}\sum_{i}\sigma_{i}y^{4}_{i}-\sum_{i}\mu_{i}y^{2}_{i} ,
\end{align}
where $y_{i}$ are linear combinations of $x_{i}$ that diagonalize $E$. The mean energy per cell is therefore approximated by
\begin{align}
\frac{\left\langle E\right\rangle}{N}=\frac{E_{0}}{N}-\frac{T}{N}\frac{\frac{\partial}{\partial\alpha}\int e^{-\frac{1}{2}\alpha\left(\sum_{i}\sigma_{i}y^{4}_{i}-\sum_{i}\mu_{i}y^{2}_{i}\right)/T}d\vec{y}}{\int e^{-\frac{1}{2}\left(\sum_{i}\sigma_{i}y^{4}_{i}-\sum_{i}\mu_{i}y^{2}_{i}\right)/T}d\vec{y}}\Big|_{\alpha=1} .
\end{align}
Changing the integration variables to $z_{i}=y^{2}_{i}$ yields
\begin{align}
\frac{\left\langle E\right\rangle}{N}=\frac{E_{0}}{N}-\frac{T}{N}\frac{\frac{\partial}{\partial\alpha}\int e^{-\frac{1}{2}\alpha\left(\sum_{i}\sigma_{i}z^{2}_{i}-\sum_{i}\mu_{i}z_{i}\right)/T}\frac{d\vec{z}}{\prod_{i}2\sqrt{z_{i}}}}{\int e^{-\frac{1}{2}\left(\sum_{i}\sigma_{i}z^{2}_{i}-\sum_{i}\mu_{i}z_{i}\right)/T}\frac{d\vec{z}}{\prod_{i}2\sqrt{z_{i}}}}\Big|_{\alpha=1} .
\end{align}
Evaluating the integrals yields
\begin{align}
\frac{\left\langle E\right\rangle}{N}=\frac{E_{0}}{N}-\frac{T}{N}\frac{\frac{\partial}{\partial\alpha}\prod_{i}e^{\alpha w_{i}}K_{\frac{1}{4}}\left(w_{i}\right)}{\prod_{i}e^{w_{i}}K_{\frac{1}{4}}\left(w_{i}\right)}\Big|_{\alpha=1} ,
\end{align}
where $K$ is the modified Bessel function of the second kind and we defined for brevity
\begin{align}
w_{i}=\frac{\mu^{2}_{i}}{4T\sigma_{i}} .
\end{align}
Performing the derivative and setting $\alpha=1$ yields
\begin{align}
\frac{\left\langle E\right\rangle}{N}=\frac{E_{0}}{N}-\frac{T}{N}\sum_{i}\left(1-\frac{K_{3/4}(w_{i})+K_{5/4}(w_{i})}{2K_{1/4}(w_{i})}\right)w_{i} .
\end{align}
For small $T$ (and thus large $w_{i}$) we use the asymptotic expansion of the modified Bessel function such that
\begin{align}
&\frac{\left\langle E\right\rangle}{N}=\frac{E_{0}}{N}-\frac{T}{N}\sum^{2n}_{i}\left(-\frac{1}{2}+\frac{3T}{8}\frac{\sigma_{i}}{\mu^{2}_{i}}\right)=\nonumber\\
&=\frac{E_{0}}{N}+\frac{n}{N}T-\frac{3T^{2}}{8}\left\langle\frac{\sigma}{\mu^{2}}\right\rangle .\label{evsd_2app}
\end{align}

\end{document}